\documentclass[%
 reprint,
nofootinbib,
 amsmath,amssymb,
 aps,
 prd,
 showkeys,
]{revtex4-2}

\usepackage{aas_macros}
\usepackage{graphicx}
\usepackage{dcolumn}
\usepackage{bm}
\usepackage{verbatim} 
\usepackage{braket}
\usepackage{hyperref}
\usepackage{mathtools}

\begin{document}

\preprint{APS/123-QED}

\title{Invariants in Co-polar Interferometry: an Abelian Gauge Theory}

\author{Nithyanandan Thyagarajan}%
\email{Nithyanandan.Thyagarajan@csiro.au}
\homepage{https://tnithyanandan.wordpress.com}
\affiliation{CSIRO, Space \& Astronomy, P. O. Box 1130, Bentley, WA 6102, Australia}%
\affiliation{National Radio Astronomy Observatory, 1003 Lopezville Rd, Socorro, NM 87801, USA.}%

\author{Rajaram Nityananda}
\email{rajaram.nityananda@gmail.com}
\affiliation{Azim Premji University, Bikkanahalli Main Road, Sarjapura, Bengaluru 562125, Karnataka, India}%

\author{Joseph Samuel}
\email{sam@rri.res.in}
\affiliation{International Centre for Theoretical Sciences, Bengaluru 560089, Karnataka, India} 
\affiliation{Raman Research Institute, Bengaluru 560080, Karnataka, India}%

\date{\today}
\begin{abstract}

An $N$-element interferometer measures correlations among pairs of array elements. Closure invariants associated with closed loops among array elements are immune to multiplicative, element-based (``local'') corruptions that occur in these measurements. Till recently, it has been unclear how a complete set of independent invariants can be analytically determined. We view the local, element-based corruptions in co-polar correlations as gauge tranformations belonging to the gauge group $\textrm{GL}(1,\mathbb{C})$. Closure quantities are then naturally gauge invariant. We use this to provide a simple and effective formalism, and identify the complete set of independent closure invariants from co-polar interferometric correlations using only quantities defined on $(N-1)(N-2)/2$ elementary and independent triangular loops. The $(N-1)(N-2)/2$ closure phases and $N(N-3)/2$ closure amplitudes (totaling $N^2-3N+1$ real invariants), familiar in astronomical interferometry, naturally emerge from this formalism, which unifies what has required separate treatments until now. We do not require auto-correlations, but can easily include them if reliably measured.
This unified view clarifies issues relating to noise and inference of object model parameters. It also allows us to extend the rule of parallel transport associated with Pancharatnam phase in optics to apply to amplitudes as well. The framework presented here extends to $\textrm{GL}(2,\mathbb{C}$) for full polarimetric interferometry as presented in a companion paper, which generalizes and clarifies earlier work. Our findings are relevant to state of the art co-polar and full polarimetric very long baseline interferometry measurements to determine features very near the event horizons of blackholes at the centers of M87, Centaurus~A, and the Milky Way.

\end{abstract}


\keywords{Gauge theories; Gauge theory techniques; Geometric \& topological phases; Geometrical \& wave optics; Group theory; Imaging \& optical processing; Interferometry; Lattice gauge theory; Mathematical physics; Radio, microwave, \& sub-mm astronomy}
\maketitle


\section{Introduction}\label{sec:intro}

Interferometry, which began with Young's double slit experiment, has blossomed over the years into a technique now widely used in physics, biology and astronomy. Examples include the use of astronomical interferometry in the first determination of the double-lobed morphology of Cygnus~A \cite{Jennison+1953}, the first imaging of the event horizon of a black hole in the center of the M87 galaxy \cite{eht19-1,eht19-2,eht19-3,eht19-4,eht19-5,eht19-6,eht21-7} and inner jet structures around the black hole in Centaurus A \cite{Janssen+2021}, 
determination of crystal structures \cite[][and references therein]{Hauptman1986,Hauptman-Nobel,Hauptman1991,Giacovazzo+2002,Giacovazzo2014}, seismic imaging \cite[][and references therein]{Snieder+2013}, remote sensing using radar and sonar \cite[][and references therein]{Callow2003,Rosen+2000}. 

In each of these applications, accurate measurement of both the amplitude and phase of the coherence (represented as complex values) is critical for success. Often, accurate measurements of the complex-valued correlations is made difficult by corruptions introduced by the propagation medium, and non-ideal behavior of the instrument as well as of the measurement process. For example, in radio interferometry, the electromagnetic wavefronts are corrupted by ionospheric and tropospheric turbulences at low and high frequencies, respectively, as well as by corrupting factors in the array element responses. In optical interferometry, the atmospheric turbulence and imperfections in the telescope's surface geometry tend to destroy the phase information. 

These element-based effects can be calibrated if there is a standard object of known morphology available \cite{TMS2017,SIRA-II}. An alternative strategy is self-calibration \cite{Cornwell+1981,Ekers1984}, which is a self-consistent scheme which uses iteration and feedback to successively refine the image starting with an initial model. But this is not always possible at the desired level of accuracy when the signal-to-noise ratio or the \textit{a priori} knowledge of the calibrator object is inadequate. Therefore, there is considerable interest in calibration-independent quantities that are unaffected by these element-specific factors, and are hence true observable properties of the object's structure. 

Such interferometric invariants are typically constructed using the product of pairwise correlations measured on a closed loop of array elements, and we call them \textit{closure invariants}. In co-polar interferometry (correlations between measurements of the same polarization at all the array elements) at radio and optical wavelengths, they are widely known as closure phases and amplitudes \cite[][and references therein]{TMS2017,SIRA-II,mon07b}. 
The analog of closure invariants include the triplet and quartet invariants in X-ray crystallography \cite[][and references therein]{Hauptman1986,Hauptman-Nobel,Hauptman1991,Giacovazzo+2002,Giacovazzo2014}, 
and Bargmann invariants in quantum mechanics \cite{Bargmann1964}. A detailed geometric insight into closure phase and its connections to various disciplines is provided in \cite{Thyagarajan+2020c}. 

In astronomical interferometry, closure quantities can be traced to \cite{jen58,Twiss+1960}. Since then, they have been invaluable tools for interferometry at optical \cite{mon03a,mon03b,mon06,mon07a,mon07b} and radio frequencies \cite[][and references therein]{Schwab1980,Cornwell+1981,Pearson+1984,bhatnagar+2001}. This is particularly true of high spatial resolution applications using very long baseline interferometry (VLBI) at radio frequencies. A famous example is the recent imaging of the event horizon of the supermassive black hole at the center of the M87 galaxy \cite{eht19-1,eht19-2,eht19-3,eht19-4,eht19-5,eht19-6,eht21-7}. Recently, new approaches of probing the structure formation in the intergalactic medium through application of closure phases on faint spectral line emission from the early universe during the cosmic reionization epoch ($z\gtrsim 6$) are also being explored \cite{thy18,thy20a,thy20b}. 

In the context of co-polar correlations in astronomical interferometry, a detailed mathematical approach to determining the number of generic closure invariants (including closure phases and closure amplitudes), and implications for the resulting signal-to-noise ratios was presented by \cite{Lannes1991}. Recently, first steps in the extension of closure invariants (called ``closure traces'') to correlations of full polarimetric antenna measurements were taken in \cite{Broderick+2020}, which relied on including antenna auto-correlations to derive an independent set of closure traces involving four correlations. 

In this and an accompanying paper \cite{polarimetric-invariants} (hereafter papers~I and II, respectively), we establish a formalism using a combination of group-theoretic and linear algebraic approaches that advances the previous work \cite{Lannes1991,Broderick+2020}. In contrast to \cite{Broderick+2020}, we do not rely on the use of auto-correlations (which in radio astronomy tend to be susceptible to significant noise biases, besides instabilities caused by radio frequency interference and instrumental systematics), but can incorporate them if reliably measured, and derive a complete and independent set of closure invariants using gauge theory. In this paper, we employ the $\mathrm{GL}(1,\mathbb{C})$ gauge group\footnote{In general, $\textrm{GL}(n,\mathbb{C})$ refers to the group of nonsingular $n\times n$ matrices under multiplication, so $\textrm{GL}(1,\mathbb{C})$ is the group of nonzero complex numbers under multiplication.}
and its associated gauge freedom to treat and derive the complete and independent set of closure invariants in the co-polar case. The closure phases and closure amplitudes familiar in radio interferometery emerge naturally from our formalism which treats triangular loops as fundamental units, and thus unifies the prescription for obtaining closure phases and closure amplitudes. In paper~II, we build on the foundations presented in this paper and provide a formalism using the $\textrm{GL}(2,\mathbb{C})$ gauge group and its associated gauge freedoms for deducing the complete and independent set of closure invariants in the general case of full polarimetric interferometry. In both papers, we also confirm our findings with a parallel and independent viewpoint using conventional linear algebra along with numerical simulations. We also generalize the analysis to provide a prescription for extracting all the independent closure invariants in an $N$-element interferometer array. These closure invariants represent true observables about the system under observation, which in the case of interferometric imaging in astronomy corresponds to the true physical properties of the target object's morphology.

This paper is structured as follows. Section~\ref{sec:context} lays out the co-polar interferometry context within which we seek a full set of independent invariants. In section~\ref{sec:counting}, we present the expected number of independent closure invariants through a dimension count analysis. Section~\ref{sec:closure-invariants} develops a formalism that provides multiple but equivalent methods to produce a complete and independent set of closure invariants, which can be readily identified with closure phases and closure amplitudes familiar in radio interferometry. These invariants are listed explicitly in the case of three, four, and $N$ array elements. Section~\ref{sec:summary} summarizes this work. In Appendix~\ref{sec:short-spacing}, we describe a scheme for reliably measuring auto-correlations as the coincidence limit of cross-correlations if an extra short-spaced pair of elements is available. In Appendix~\ref{sec:selfcal}, we place our work in relation to the widely used `self-calibration' technique, which implicitly uses closure quantities. Appendix~\ref{sec:numerics} describes the numerical scheme used to confirm our analytic results, and Appendix~\ref{sec:invariants-form} brings out how our approach clarifies some aspects of the choice of invarints as it relates to the effects of noise and to object model parameters. In Appendix~\ref{sec:Pancharatnam-closure}, we highlight the explicit connection between the closure phase and Pancharatnam phase \cite{Pancharatnam1956a,Pancharatnam1975} in optics. Motivated by the radio astronomy application, we extend the Pancharatnam  rule to include amplitudes as well.

\section{Co-polar interferometry}\label{sec:context}

Consider an interferometer array with $N$ elements labeled by the indices $a,b=0,\ldots,N-1$. Each element, $a$, in the array measures the amplitude and phase of the same polarization (co-polar, or scalar) state of electric field, $e_a$ (represented by a complex number), incident on it, which is stochastic. The true correlation of the stochastic fields between pairs of array elements is, by definition, obtained by cross-multiplication and averaging, $S_{ab}\coloneqq \left\langle e_a e_b^\dagger \right\rangle$, where, $\dagger$ denotes the conjugate transpose operator, which reduces to complex conjugation for $\textrm{GL}(1,\mathbb{C})$ matrices in the co-polar interferometric case dealt in this paper. 

The complex electric fields, $e_a$, may be subject to arbitrary gains $G_a$ at each element, where, $G_a$ are non-zero complex numbers representing both amplitude and phase distortion of the measured signal. In ideal conditions, all the gains $G_a$ would be unity and the measured signal at each baseline would accurately reflect the true signal correlation, $S_{ab}$, of emission from the object. In fact, the measured correlation, $C_{ab}$, which is a corrupted form of the true correlation, can be written as
\begin{align}
    C_{ab} &= G_a \left\langle e_a e_b^\dagger \right\rangle G_b^\dagger = G_a \, S_{ab} \, G^\dagger_b \label{mtrelation}
\end{align}
because of local gain distortions at each element. Generally, the cross-correlations are complex-valued with non-zero amplitudes. From the definition, cross-correlations satisfy $C_{ab}=C_{ba}^\dagger$. The auto-correlations, $A_{aa}\coloneqq C_{aa}$, are real and positive. Our objective is to construct quantities which are immune to the gain distortions and actually reflect true properties of the object's morphology, rather than local conditions at the individual elements of the interferometer. 

We relate this problem to the `gauge' theories of fundamental forces like electromagnetism. The local gains $G_a$ are regarded as gauge transformations and our objective is to isolate a maximal set of gauge invariant quantities which we will call closure invariants. This term encompasses both closure amplitudes and closure phases. Our treatment places them on an equal footing by working with a $\textrm{GL}(1,\mathbb{C})$ gauge group.

\section{Counting arguments}\label{sec:counting}

Let us first do a dimensional count to see how many closure invariants we would expect to find. The number of measured cross-correlations is $N(N-1)/2$, this being the number of baselines or element pairs, i.e., the number of non-repeating combinations among $N$ elements taken two at a time. Since each cross-correlation is a complex number we have $N(N-1)$ real numbers. If $n_\textrm{A}$ auto-correlations (described by $n_\textrm{A}$ real numbers) are also measured, we would have to add $n_\textrm{A}$ to the above count. Without loss of generality, we can assume that $n_\textrm{A}$ is either zero or one\footnote{In radio astronomy, measuring any more auto-correlations does not give us new information due to the implicit assumption of spatial stationarity.}. In radio astronomy, auto-correlation measurements are unreliable because they are dominated by non-astronomical systematics. Our formalism for closure invariants works with or without auto-correlations. 

The unknown element-based gains are $N$ complex numbers, but note that there may be sets of gains $G_a$ which do not affect 
the correlations. That is, they satisfy $G_a C_{ab}G_b^\dagger=C_{ab}$ for all $a,b$. This yields $G_a G_b^\dagger=1$ and so $G_aG_a^\dagger G_b G_b^\dagger=1$. If the correlations themselves are unchanged, then so will the triple products of correlations. Thus, using $G_a G_b^\dagger G_b G_c^\dagger G_c G_a^\dagger=1$ (assuming $N\ge 3$, and distinct $a,b,c$) gives us $G_c G_c^\dagger=1$ for all $c$. This means $G_a=e^{i\theta}$ for all $a$, which is simply an overall phase factor that cancels out in Eq.~(\ref{mtrelation}). This gives us a count of $2N-1$ real numbers in the unknown gains.

Assuming none of the measured correlations is redundant (they are all independent of each other) and all of them are used, the number of closure invariants (that are independent of any choice of the unknown gains) is the difference
\begin{align}
  N(N-1)+n_\textrm{A} - (2N-1) &= N^2-3N+1+n_\textrm{A} \, . \label{eqn:count}
\end{align}
Note that this equals the number of closure phases, $(N-1)(N-2)/2$, plus the number of closure amplitudes, $N(N-3)/2+n_\textrm{A}$ \cite{TMS2017,Lannes1991}.
For example, this equals 1 (one closure phase) or 5 (three closure phases and two closure amplitudes) when $n_\textrm{A}=0$ for $N=3$ or $N=4$, respectively, and 2 (one closure phase and one closure amplitude) or 6 (three closure phases and three closure amplitudes) when $n_\textrm{A}=1$ for $N=3$ and $N=4$, respectively. As $N$ becomes large, the structural information about the object that can be extracted using closure phases and closure amplitudes asymptotically approaches that which can be extracted using correlations.

\section{Formalism for closure invariants}\label{sec:closure-invariants}

From a gauge theory perspective, we can regard each of the $N$ elements $a=0,\ldots,N-1$ as vertices in a graph. The gains that represent element-based corruption factors are local variables and multiplication by the gain factor results in a gauge transformation at each vertex. Each baseline is an edge $ab$ or link carrying ``connection'' variables that are bilocal since they are defined on the link connecting two vertices. These are the cross-correlations. In the language of gauge theory, each triangle, $\Delta_{(abc)}\equiv (a,b,c)$ is called an ``elementary plaquette'' and defines a ``Wilson loop''. The Wilson loop of any closed circuit can be determined from those of the elementary triangular plaquettes. Our objective is to list a complete and independent set of elementary Wilson loops. These are the closure invariants of radio astronomy.

In a $U(1)$ gauge theory like electromagnetism, the gauge group would be unitary and complex conjugation in Eq. (\ref{mtrelation}) would be the same as inversion. One can then form independent closure invariants from each elementary triangular plaquette $abc$ by simply multiplying correlations around the triangle, as in $\mathcal{B}_{abc}=C_{ab}C_{bc}C_{ca}$. ($\mathcal{B}_{abc}$ is called the ``bispectrum'' in interferometry). However, our gauge group $\textrm{GL}(1,\mathbb{C})$ is {\it not} unitary. 
Using $\mathcal{G}=\{G_a\}, a=0,\ldots,N-1$ to denote a general gauge transformation at every vertex, $a$, on the graph, the bispectrum and other higher order products (``polyspectra'') of a set of $C_{ab}$ around a closed loop, $\Gamma$, transform as 
\begin{align}
  \mathcal{P}_\Gamma &\xmapsto{\mathcal{G}} |G_a|^2 \, |G_b|^2 \ldots \mathcal{P}_\Gamma \, .
\end{align}
The phase of $\mathcal{P}_\Gamma$ is invariant under gauge transformations $\mathcal{G}=\{G_a\},a=0,\ldots,N-1$ and this is familiarly known as the closure phase. However, because the gauge group is not unitary, the amplitude of $\mathcal{P}_\Gamma$ is not invariant as it still depends on the gain amplitudes, $|G_a|$.

Similarly, one can define closure amplitudes for every even-edged loop. For example, given the loop $\Gamma=abcda$, the quantity
\begin{align}
  \mathcal{Q}_{\Gamma} &= C_{ab} \, C^{-1}_{bc} \, C_{cd} \, C^{-1}_{da}
\end{align}
transforms as 
\begin{align}
  \mathcal{Q}_\Gamma &\xmapsto{\mathcal{G}} G_a G^{\dagger -1}_a \, G_b G^{\dagger -1}_b \,G_c G^{\dagger -1}_c \, G_d G^{\dagger -1}_d \mathcal{Q}_\Gamma .
\end{align}
It follows, since $\left| G_a G^{\dagger -1}_a \right|=1$ that $\left|\mathcal{Q}_\Gamma\right|$, the modulus of $\mathcal{Q}_\Gamma$, is a closure invariant:
\begin{align}
  \left|\mathcal{Q}_\Gamma\right| &\xmapsto{\mathcal{G}} \left|\mathcal{Q}_\Gamma\right| .
\end{align}
However, the phase of $\mathcal{Q}$ is {\it not} an invariant because it still depends on the phases of $G_a$ at the loop vertices.

The virtues of $\mathcal{P}_\Gamma$ and $\mathcal{Q}_\Gamma$ can be combined in a single quantity $\mathcal{C}_\Gamma$ by defining a hat operator, $\widehat{Z}=(Z^\dagger)^{-1}$, on non-zero complex numbers. We introduce the term \textit{covariants} as the set of even number of products of correlations around a closed loop, $\Gamma$, with even numbered terms ``hatted'' starting with the second. For example, a covariant on a 4-vertex loop can be written as $\mathcal{C}_\Gamma = C_{ab}\widehat{C}_{bc}C_{cd}\widehat{C}_{da}$. Hence, covariants transform as
\begin{align}
  \mathcal{C}_\Gamma &\xmapsto{\mathcal{G}} G_0 \, \mathcal{C}_\Gamma \, G_0^{-1} \, .
\end{align}
Because the gauge group of $G_a$, $\textrm{GL}(1,\mathbb{C})$, is Abelian, 
\begin{align}
  \mathcal{C}_\Gamma &\xmapsto{\mathcal{G}} \mathcal{C}_\Gamma \, .
\end{align}
Thus, the covariants are invariant\footnote{Note that a four-edged covariant essentially resembles a \textit{cross-ratio}, which is a well-known invariant in projective geometry \cite[][and references therein]{Klein1872,Penrose+1984}.} under the gauge transformation effected by the element-based corruption, and form the closure invariants we are seeking in co-polar interferometry. Some of the notation above is deliberately kept general so that it transfers easily to the more difficult non-Abelian case discussed in paper~II. 

\subsection{Triangular Plaquettes}\label{sec:triangular-plaquettes}

The need to have an even number of correlations in the product around a closed loop appears to impose the requirement of elementary quadrilaterals rather than triangles. However, it is unclear beforehand which even-vertexed loops will provide us with a complete set of independent invariants. We show below that such a set can be obtained by considering variables on elementary triangular plaquettes that are independent, rather than quadrilateral plaquettes. The latter can be entirely derived from the former.

The independent triangles in an $N$-element interferometer can be obtained by fixing a base vertex (for example, 0) and choosing all the triangles that contain this vertex \cite{TMS2017}. There are $N_\Delta=(N-1)(N-2)/2$ such independent triangles. We construct elementary triangular plaquette variables pinned at vertex $0$ that includes just three elements ($0, a, b$): $\mathcal{A}_{ab}=C_{0a}\widehat{C}_{ab}C_{b0}$. This quantity is neither a bispectrum nor a covariant, but acts as a building block that can be used to construct all the closure invariants. These triangular variables are clearly independent and complete as all closed loops can be decomposed into triangles and each distinct triangle is independent. Around the closed loop pinned at vertex 0 ($\Gamma_0$), they undergo gauge transformations with $G_0$ as
\begin{align}
\mathcal{A}_{ab} &\xmapsto{\mathcal{G}} \left|G_0\right|^2\mathcal{A}_{ab} \, . \label{eqn:Atransform}
\end{align}
We term such quantities as {\it advariants} in this paper. A special advariant is $\mathcal{A}_0=A_{00}$, the auto-correlation. The notation clarifies that this auto-correlation advariant is a local variable based at $0$. Note that advariants have an advantage that their gauge freedom is restricted to just a scaling by a single $|G_0|^2$ in contrast to polyspectra (including bispectrum) whose gauge transformations depend on $|G_a|$ of all the vertices in the loop. Thus, $|G_0|^2$ is the only unknown quantity in the advariants that has to be eliminated to arrive at invariant quantities. However, the phase of a triangular advariant is equal to the phase of the corresponding bispectrum, and thus gives the closure phase \cite{jen58}, which is an invariant. 

\subsection{Complete and Independent Set of Invariants}\label{sec:invariants}

\subsubsection{Method 1}\label{sec:method1}

Consider all the $N_\Delta$ elementary triangular plaquettes pinned at vertex $0$. Each of them gives us a complex number $\mathcal{A}_{ab}$. These complex numbers scale as in Eq.~(\ref{eqn:Atransform}). The real and imaginary parts of these complex numbers give us $M=2N_\Delta$ real numbers, $\{\widetilde{I}_m\}, m=1,\ldots,M$, all of which transform by the unknown scale, $|G_0|^2$, under gauge transformations. The {\it ratios} of these $\widetilde{I}_m$ with respect to any of them ($\widetilde{I}_0$, for example) or some symmetric combination ($[\sum_m \widetilde{I}_m^2]^{1/2}$, for example), will eliminate this unknown scale, and thus give us $M-1$ real closure invariants, $\{\widetilde{I}_k/\widetilde{I}_0\}, k=2,\ldots,M$. Their number is $N^2-3N+1$, which is in exact agreement with the expectation in Eq.~(\ref{eqn:count}). This gives us a complete and independent set of closure invariants. Also, there is no need for auto-correlations in this construction. If we were to add $A_{00}$ as a measured quantity, we would gain a single real invariant because $\mathcal{A}_0=A_{00}$ being real and positive, would only contribute one more real number to $\{\widetilde{I}_m\}$. Note that the practical problem of systematic effects affecting auto-correlations can be mitigated by the use of cross-correlations between a pair of closely spaced elements in cases where the angular size of the object is much smaller than the angular resolution corresponding to the short spacing, and is described in Appendix~\ref{sec:short-spacing}. 

While this method of taking ratios completely solves the problem we sought to address, namely, the determination of a complete and independent set of co-polar interferometric invariants, we provide below an
alternative formalism that allows us to better visualize the closure phases and amplitudes, as they are familiarly known in radio interferometry, in terms of combinations of the invariants determined here. 

\subsubsection{Method 2}\label{sec:method2}

The elementary covariant quantity defined earlier as the closed loop product of an even number of correlations with alternate terms hatted starting with the second (a minimum of 4 terms) can be equivalently defined as the multiplication of an even number (minimum of two) advariants with every alternate term hatted starting with the second. For example, consider two triangles, $\Delta_1\equiv \Delta_{(0,1,2)}$ (the base triangle) and $\Delta_\ell\equiv \Delta_{(0,a,b)}$ chosen from the set of $N_\Delta$ triangles, both pinned at vertex 0, and their reversed forms, $\nabla_1\equiv \Delta_{(0,2,1)}$ and $\nabla_\ell\equiv \Delta_{(0,b,a)}$. 
The covariant, which is an invariant, can now be expressed by pairing the advariants on these triangles as 
\begin{align}
  \mathcal{I}_{\Delta_1;\Delta_\ell} = \mathcal{C}_{\Delta_1;\Delta_\ell} &= \mathcal{A}_{\Delta_1}\widehat{\mathcal{A}}_{\Delta_\ell} \label{eqn:rect-covariant} \\
  \textrm{or,} \quad \mathcal{I}_{\Delta_1;0} = \mathcal{C}_{\Delta_1;0} &= \mathcal{A}_{\Delta_1}\widehat{\mathcal{A}}_0 \, ,\label{eqn:tri-covariant}
\end{align}
which are in general complex-valued. In the co-polar case studied in this paper, since covariants are indeed invariants, they will be used interchangeably. However, the same is not true in the full polarimetric case presented in paper~II. 

The following properties of covariants are noted:
\begin{itemize}
  \item $\mathcal{A}_{\Delta_1}$ scales as $|G_0|^2$, whereas $\widehat{\mathcal{A}}_{\Delta_\ell}$, and $\widehat{\mathcal{A}}_0$ scale as $|G_0|^{-2}$, thereby cancelling the unknown scale factor in Eqs.~(\ref{eqn:rect-covariant}) and (\ref{eqn:tri-covariant}), and making the covariant an invariant.
  \item A covariant formed with $\Delta_1$ and $\Delta_\ell$ has the same amplitude as that formed with $\Delta_1$ and $\nabla_\ell$. That is, $|\mathcal{I}_{\Delta_1;\Delta_\ell}| = |\mathcal{I}_{\Delta_1;\nabla_\ell}|$, and thus do not provide independent amplitude information. Similarly, $|\mathcal{I}_{\Delta_\ell;\Delta_1}| = |\mathcal{I}_{\Delta_1;\Delta_\ell}|^{-1}$.
  \item If $\phi_{\Delta_1}$ and $\phi_{\Delta_\ell}$ are the phases of the advariants (closure phases), then the phases of $\mathcal{I}_{\Delta_1;\Delta_\ell}$ and $\mathcal{I}_{\Delta_1;\nabla_\ell}$ are $\phi_{\Delta_1}+\phi_{\Delta_\ell}$ and $\phi_{\Delta_1} - \phi_{\Delta_\ell}$, respectively. 
  \item Each invariant, $\mathcal{I}_{\Delta_1;\Delta_\ell}$, is complex-valued and thus contains two real-valued invariants, except when $\Delta_\ell=\Delta_1$. In that case, $|\mathcal{I}_{\Delta_1;\Delta_1}|=1$, and therefore, contains only one real invariant, namely, its phase which is twice the closure phase, $2\phi_{\Delta_1}$.
\end{itemize}

Using the above properties, a set of covariants (invariants) can be constructed by pairing all the independent triangular advariants, $\mathcal{A}_{\Delta_\ell}$ with the advariant of the base triangle, $\mathcal{A}_{\Delta_1}$, as $\mathcal{I}_{\Delta_1\Delta_\ell}=\mathcal{A}_{\Delta_1}\widehat{\mathcal{A}}_{\Delta_\ell}, \ell=1,\ldots,N_\Delta$. Each of these complex invariants yields two real invariants for $\ell\ne 1$ and only one real invariant for $\ell=1$ as noted above. Therefore, the total number of real invariants is $2N_\Delta-1 = N^2-3N+1$. The presence of an auto-correlation measurement will increase this count to $N^2-3N+2$ due to one additional real-valued amplitude invariant in $\mathcal{A}_{\Delta_1}\widehat{\mathcal{A}}_0$. Therefore, this method provides $N^2-3N+1+n_\textrm{A}$ real invariants, which is a complete and independent set, and is consistent with the first method using ratios as well as with the dimension count analysis in section~\ref{sec:counting}. We also verified these results numerically as detailed in Appendix~\ref{sec:numerics}.

\subsection{Closure Phases and Amplitudes}\label{sec:closure-phase-amp}

Using the second method, we can directly identify the closure phases and amplitudes familiar in radio interferometry with the closure invariants obtained here. 

The independent closure phases are directly given by the phases of the $N_\Delta$ independent triangular advariants. But they can be derived from the covariants, $\mathcal{I}_{\Delta_1\Delta_\ell}$, as well, where $\Delta_\ell$ is any triangle that shares vertex 0 with the reference triangle, $\Delta_1$. The phase of $\mathcal{I}_{\Delta_1\Delta_1}$ is $2\phi_{\Delta_1}$, which yields $\phi_{\Delta_1}$ to within an ambiguity of $\pi$. This ambiguity can be addressed by using the phase of the advariant, $\mathcal{A}_{\Delta_1}$, which gives $\phi_{\Delta_1}$ directly without the aforementioned ambiguity. The phase of the rest of the covariants $\mathcal{I}_{\Delta_1\Delta_\ell}$ is $\phi_{\Delta_1}+\phi_{\Delta_\ell}$, which will directly yield $\phi_{\Delta_\ell}$ via plain substitution of the previously determined $\phi_{\Delta_1}$. This will yield $N_\Delta$ independent closure phases. The same construct also yields closure amplitudes.

An inspection of $|\mathcal{I}_{\Delta_1\Delta_\ell}|$ shows that it is indeed a closure amplitude, $|C_{01}| |C_{21}|^{-1} |C_{20}| |C_{a0}|^{-1} |C_{ab}| |C_{0b}|^{-1}$. In general, closure amplitudes formed from a pair of triangular advariants involves six such terms. However, when $\Delta_\ell$ shares an edge with $\Delta_1$ ($a=2$, for example), then the contributions, $|C_{20}|$ and $|C_{a0}|^{-1}$, associated with that edge cancel each other, and the closure amplitude reduces to the familiar form containing four terms, $|C_{01}| |C_{21}|^{-1} |C_{2b}| |C_{0b}|^{-1}$. Noting that $|\mathcal{I}_{\Delta_1\Delta_1}|$ does not contribute an amplitude (it always has unit amplitude), there are $N_\Delta-1=N(N-3)/2$ independent closure amplitudes (from the $N_\Delta-1$ triangle pairs sharing vertex 0 with the base triangle $\Delta_1$). The auto-correlation measurement, if present, adds one more independent closure amplitude through $|\mathcal{A}_{\Delta_1}\widehat{\mathcal{A}}_0|=|C_{01}| |C_{21}|^{-1} |C_{20}| A_{00}^{-1}$. Therefore, there are $N_\Delta-1+n_\textrm{A}$ independent closure amplitudes in general. It may also be noted that if an auto-correlation can be reliably measured, we can obtain all the invariants just from triangles with the auto-correlation advariant as the base advariant using $\mathcal{I}_{0;\Delta_l} = \mathcal{A}_0\widehat{\mathcal{A}}_{\Delta_1}$, without requiring quadrilaterals or 6-edged loops.

The set of closure amplitudes and phases contains a total of $2N_\Delta-1+n_\textrm{A}=N^2-3N+1+n_\textrm{A}$ real invariants as given in Eq.~(\ref{eqn:count}), which is complete and independent, and is in agreement with all the earlier analyses. Our approach using advariants and covariants (invariants) constructed only from these triangular plaquettes provides a unifying approach that directly identifies the traditional invariants known as closure phases and amplitudes in radio interferometry. 

Then the question arises: which form of closure invariants is preferable? In Appendix~\ref{sec:invariants-form}, we consider this question from the viewpoint of correlations within the chosen set of invariants and their overall contribution to the determination of the likelihood of parameterized models. In brief, the choice of invariants is immaterial in the determination of the likelihood function of the model parameters as long as the set of invariants employed is complete. However, certain choices of coordinates in which the invariants are represented could lead to coordinate-induced correlations among the invariants, and also to situations when one or more of the invariants become ill-defined in those coordinates. A well-known example occurs when representing real and imaginary parts of random variable in polar form, where the amplitude and phase angle can become correlated in general and the phase angle becomes ill-defined in low $S/N$ regimes even if the real and imaginary parts are perfectly uncorrelated and well-defined. For these reasons, even if the real and imaginary parts of the invariants in our approach may be correlated, we still prefer this representation over the polar form consisting of phases and amplitudes.

\subsection{Illustration with Examples}\label{sec:examples}

For illustration, we consider the familiar examples of 3- and 4-elements arrays below.

\subsubsection{3-element Array}

Consider an array with $N=3$ and one auto-correlation measured ($n_A=1$). We have 1 and 6 real values in the auto- and cross-correlations, respectively, totaling 7 real measurements. We have $2N-1=5$ unknown real-valued parameters in the three complex-valued element gains. From dimension counting used in Eq.~(\ref{eqn:count}), we expect to find two real-valued closure invariants. From these measurements, we can form a triangular advariant, $\mathcal{A}_{\Delta_1}$, corresponding to the triangle $\Delta_1\equiv \Delta_{(0,1,2)}$, and an auto-correlation advariant, $\mathcal{A}_0$. The complex invariant 
\begin{align}
  \mathcal{I}_{\Delta_1;0} &= \mathcal{A}_{\Delta_1}\widehat{\mathcal{A}}_0 = \frac{|C_{01}| |C_{20}| }{|C_{12}| A_{00}} \, e^{i\phi_{\Delta_1}} \label{eqn:auto-corr-covariant}
\end{align}
gives both these real-valued invariants. The magnitude of this complex quantity is like a closure amplitude (with one auto-correlation) and the phase is the standard closure phase of three elements. These two quantities are evidently independent. $\mathcal{I}_{\Delta_1;\Delta_1}=e^{2i\phi_{\Delta_1}}$ will give no new invariants.

If auto-correlation is not measured ($n_A=0$), the number of real measurements reduces by 1 to 6. Hence, we now expect only one invariant from Eq~(\ref{eqn:count}). Using the only advariant available, $\mathcal{A}_{\Delta_1}$, we find 
\begin{align}\label{three}
  \mathcal{I}_{\Delta_1;\Delta_1} &= \mathcal{A}_{\Delta_1}\widehat{\mathcal{A}}_{\Delta_1} = e^{2i\phi_{\Delta_1}} \, 
\end{align}
contains that expected invariant, namely, its phase, which is twice the standard closure phase associated with $\Delta_1$, and carries an ambiguity of $\pi$. However, the closure phase, $\phi_{\Delta_1}$, can be unambiguously obtained from the phase of $\mathcal{A}_{\Delta_1}$ directly. 

\subsubsection{4-element Array}

For an $N=4$ array, we have $N_\Delta=3$ independent triangles, $\Delta_1\equiv \Delta_{(0,1,2)}, \Delta_2\equiv \Delta_{(0,2,3)}, \Delta_3\equiv \Delta_{(0,1,3)}$, and hence, three complex advariants, $\mathcal{A}_{\Delta_1},\mathcal{A}_{\Delta_2}$, and $\mathcal{A}_{\Delta_3}$. If the auto-correlation $\mathcal{A}_0$ is measured we have an additional real advariant. The dimension count analysis predicts six invariants when the auto-correlation is measured and five when it is not. 

When auto-correlation is not measured ($n_\textrm{A}=0$), we combine the reference triangle $\Delta_1$, first with itself, and then with the two other independent ones, $\Delta_2$ and $\Delta_3$. 
\begin{align}
  \mathcal{I}_{\Delta_1;\Delta_1} &= \mathcal{A}_{\Delta_1} \widehat{\mathcal{A}}_{\Delta_1} = e^{2i\phi_{\Delta_1}} \\
  \mathcal{I}_{\Delta_1;\Delta_2} &= \mathcal{A}_{\Delta_1} \widehat{\mathcal{A}}_{\Delta_2} = \frac{|C_{01}||C_{23}|}{|C_{21}||C_{03}|} e^{i(\phi_{\Delta_1}+\phi_{\Delta_2})} \\
  \mathcal{I}_{\Delta_1;\Delta_3} &= \mathcal{A}_{\Delta_1} \widehat{\mathcal{A}}_{\Delta_3} = \frac{|C_{20}||C_{31}|}{|C_{21}||C_{30}|} e^{i(\phi_{\Delta_1}+\phi_{\Delta_3})} \, .
\end{align}
We see that there are five real-valued invariants in the equations above -- three independent closure phases and two independent closure amplitudes familiar in radio interferometry for a 4-element co-polar interferometer array.

When $n_\textrm{A}=1$, we can write down the following three complex invariants -- $\mathcal{A}_{\Delta_1} \widehat{\mathcal{A}}_0$, $\mathcal{A}_{\Delta_2} \widehat{\mathcal{A}}_0$, and $\mathcal{A}_{\Delta_3} \widehat{\mathcal{A}}_0$ --
which gives us six real invariants comprising of three closure phases and three closure amplitudes (two as above, and one more due to auto-correlation). The invariant phase in $\mathcal{A}_{\Delta_1} \widehat{\mathcal{A}}_{\Delta_1}$ is redundant due to the presence of $\mathcal{A}_{\Delta_1} \widehat{\mathcal{A}}_0$, and can therefore be ignored. 

Thus, the closure phases and closure amplitudes emerge naturally in our approach which treats independent triangles as fundamental. We break this up into a single invariant associated with the reference triangle, $\Delta_1$, and two for each of the rest. The auto-correlation measurement will provide one additional closure amplitude. 

\subsection{Quadrilateral Closure Amplitudes}\label{sec:quads}

We have so far emphasized the role of triangles in constructing advariants. The complete set of closure invariants we generate include four-sided loops (quadrilaterals) as well as six-sided loops. Traditionally, closure amplitudes have been discussed by astronomers using quadrilaterals. In order to connect with this discussion we ask: Can we obtain a complete, independent set with only quadrilateral-based closure amplitudes? A clear methodology is provided in \cite{Blackburn+2020} which involves placing the $N$ elements on a ring and considering two non-overlapping pairs of elements where each pair is made of consecutive elements. Here, we provide an alternate method based on our formalism.

One way to ensure independence is to ensure that each loop contains a unique element pair not present in any other. To do this, we fix the element pair $(0,1)$ and consider the quadrilateral made with two more elements $a$ and $b$. We arrange for $1<a<b<N$ and choose the set (which we call $\mathbb{A}$) of closure amplitudes, $|C_{01}||C_{1a}|^{-1}|C_{ab}||C_{b0}|^{-1}$. So, each such choice of a pair $(a,b)$ with $a<b$ gives us an independent closure amplitude. The number of closure amplitudes in the set $\mathbb{A}$ is therefore the number of such $(a,b)$ pairs, which is $(N-2)(N-3)/2$. However, we know from basic counting \cite{TMS2017} and linear algebra \cite{Lannes1991} that the total number of independent closure amplitudes is $N(N-3)/2$. We therefore need $(N-3)[N-(N-2)]/2=N-3$ more independent closure amplitudes to complete the desired list. We supply these in a set $\mathbb{B}$, by observing that element $0$ is only paired with $b$ and hence never paired with element $2$ in the set $\mathbb{A}$. We therefore add precisely $N-3$ more closure amplitudes by including the pair, $(0,2)$, as $|C_{01}||C_{1b}|^{-1}|C_{b2}||C_{20}|^{-1}$, which contains $C_{02}$ in addition to our base choice of $C_{01}$. The set $\mathbb{B}$ is only indexed by $b$ which can take $N-3$ values, and its members are clearly independent of each other, since each contains the element pair $(1,a)$ which is unique to it within this set $\mathbb{A}$. Every member of $\mathbb{B}$ is also independent of all those of set $\mathbb{A}$, since none of those contains the pair $(0,2)$. This completes the explicit construction of $N(N-3)/2$ independent quadrilateral closure amplitudes. Although this construction has been described in a self-contained way, it was arrived at by considering quadrilateral closure amplitudes as the product of two advariants, sharing an edge. 

The closure amplitudes from our approach and the ring-based approach in \cite{Blackburn+2020} are essentially equivalent. However, a consequence of the fundamental difference in the methodologies leads to one notable difference in the invariants constituting the complete and independent set when an $N+1$-th element (indexed by $a=N$) is added to the array. In the approach of \cite{Blackburn+2020}, the inclusion of a new member, $a=N$, in the ring will remove the earlier pairing of $(N-1,0)$ because they are no longer adjacent and the corresponding second set of consecutive pairs will also get removed. In their place, two new pairs of consecutive indices, $(N-1,N)$ and $(N,0)$, and the corresponding sets of second adjacent pairings will have to be included. In other words, some members of the original set of closure amplitudes will be removed and replaced with a larger but different set. A similar removal and replacement process also occurs when an array element has to be excluded. In our approach, the additional element yields $N-1$ triangles (and advariants) pinned at base vertex 0 containing this element that were not present before. When paired with the base triangle, we get $N-1$ new complex-valued invariants in addition to the existing set. That is, the members of the original set are undisturbed and $2N-2$ real-valued invariants are added, amounting to a total of $N^2-N-1$ real invariants for the $N+1$-element array. This difference may have practical consequences when certain array elements are flagged at specific times or frequencies due to poor quality of data. Our approach will be robust to such practical challenges.

Thus, we have provided multiple methods of explicitly listing the closure invariants in an $N$-element interferometer array. These methods do not yield the same set of invariants, but carry exactly equivalent information. Our approach unifies the treatment of all co-polar closure invariants, including the triangular closure phases and quadrilateral closure amplitudes familiar in radio interferometry as seen earlier.

\section{Summary}\label{sec:summary}

In this paper, we have explicitly enumerated from co-polar interferometric measurements a complete and independent set of closure invariants that are immune to corruptions local to the elements of an interferometer array. The solution relies on mathematical ideas borrowed from gauge theories of particle physics and the geometric phase of quantum mechanics and optics. Specifically, the co-polar interferometric invariants presented here are invariant under gauge transformations by the element-based corruption factors which belong to the Abelian gauge group, $\textrm{GL}(1,\mathbb{C})$. 

Invariants such as closure phases and closure amplitudes have been in wide use in many disciplines for decades. However, they relied on using closed triangular and quadrilateral loops, respectively, and thus required different treatments.
The solution presented here relies on using only independent triangular plaquettes, which form the simplest non-trivial loops and are well determined, as our basis. The quantity defined on these triangles, called advariants, can be combined with an advariant chosen on a reference triangle to produce a complete set of independent invariants. The main result of this work is that using this formalism we have unified the treatment of closure invariants, which have historically required separate methods. Thus, we have provided an altered and unified perspective which improves our global understanding of interferometric closure invariants from a symmetry viewpoint. The familiar closure phases and closure amplitudes emerge naturally from this approach.

Radio astronomers are aware that auto-correlation measurements are unreliable because they could be dominated by non-astrophysical systematic effects. Our formalism works even without auto-correlations, but can naturally accommodate them if they are reliably measured. We have also outlined a methodology for using a cross-correlation on a short-spaced element pair in place of an auto-correlation to increase the number of invariants by 1, without introducing systematic errors associated with a direct single-element auto-correlation measurement.

Our approach clarifies certain aspects related to the choice of the form of closure invariants and their implications for their covariance properties and their impact on the likelihood of model parameters. While the methods presented here are specific to co-polar interferometry, the concepts serve as a stepping stone and reveal their full power when applied to a discussion of invariants in full polarimetric interferometry in paper~II. We expect that this solution will aid astronomers in processing data collected from VLBI measurements and other radio interferometry experiments.

There is also an unexpected spinoff from this application which is described in more detail in Appendix E. As explained there, there is a strong 
connection between closure phases in astronomy and the Pancharatnam phase in optics and quantum mechanics. The present application suggests how
the Pancharatnam phase rule can be generalized to include amplitudes as well as phases. This rule goes beyond existing 
ideas in the physics literature and shows how an astronomical application can enrich the physics which is being applied.

\begin{acknowledgments}
We thank Supurna Sinha, Ron Ekers, and the anonymous reviewers for a careful reading of the paper. J.~S. acknowledges support by a grant from the Simons Foundation (677895, R.G.).
\end{acknowledgments}


\begin{thebibliography}{49}%
\makeatletter
\providecommand \@ifxundefined [1]{%
 \@ifx{#1\undefined}
}%
\providecommand \@ifnum [1]{%
 \ifnum #1\expandafter \@firstoftwo
 \else \expandafter \@secondoftwo
 \fi
}%
\providecommand \@ifx [1]{%
 \ifx #1\expandafter \@firstoftwo
 \else \expandafter \@secondoftwo
 \fi
}%
\providecommand \natexlab [1]{#1}%
\providecommand \enquote  [1]{``#1''}%
\providecommand \bibnamefont  [1]{#1}%
\providecommand \bibfnamefont [1]{#1}%
\providecommand \citenamefont [1]{#1}%
\providecommand \href@noop [0]{\@secondoftwo}%
\providecommand \href [0]{\begingroup \@sanitize@url \@href}%
\providecommand \@href[1]{\@@startlink{#1}\@@href}%
\providecommand \@@href[1]{\endgroup#1\@@endlink}%
\providecommand \@sanitize@url [0]{\catcode `\\12\catcode `\$12\catcode
  `\&12\catcode `\#12\catcode `\^12\catcode `\_12\catcode `\%12\relax}%
\providecommand \@@startlink[1]{}%
\providecommand \@@endlink[0]{}%
\providecommand \url  [0]{\begingroup\@sanitize@url \@url }%
\providecommand \@url [1]{\endgroup\@href {#1}{\urlprefix }}%
\providecommand \urlprefix  [0]{URL }%
\providecommand \Eprint [0]{\href }%
\providecommand \doibase [0]{https://doi.org/}%
\providecommand \selectlanguage [0]{\@gobble}%
\providecommand \bibinfo  [0]{\@secondoftwo}%
\providecommand \bibfield  [0]{\@secondoftwo}%
\providecommand \translation [1]{[#1]}%
\providecommand \BibitemOpen [0]{}%
\providecommand \bibitemStop [0]{}%
\providecommand \bibitemNoStop [0]{.\EOS\space}%
\providecommand \EOS [0]{\spacefactor3000\relax}%
\providecommand \BibitemShut  [1]{\csname bibitem#1\endcsname}%
\let\auto@bib@innerbib\@empty
\bibitem [{\citenamefont {{Jennison}}\ and\ \citenamefont {{Das
  Gupta}}(1953)}]{Jennison+1953}%
  \BibitemOpen
  \bibfield  {author} {\bibinfo {author} {\bibfnamefont {R.~C.}\ \bibnamefont
  {{Jennison}}}\ and\ \bibinfo {author} {\bibfnamefont {M.~K.}\ \bibnamefont
  {{Das Gupta}}},\ }\bibfield  {title} {\bibinfo {title} {{Fine Structure of
  the Extra-terrestrial Radio Source Cygnus I}},\ }\href
  {https://doi.org/10.1038/172996a0} {\bibfield  {journal} {\bibinfo  {journal}
  {\nat}\ }\textbf {\bibinfo {volume} {172}},\ \bibinfo {pages} {996} (\bibinfo
  {year} {1953})}\BibitemShut {NoStop}%
\bibitem [{\citenamefont {{Event Horizon Telescope Collaboration}}\ \emph
  {et~al.}(2019{\natexlab{a}})\citenamefont {{Event Horizon Telescope
  Collaboration}}, \citenamefont {{Akiyama}}, \citenamefont {{Alberdi}},
  \citenamefont {{Alef}}, \citenamefont {{Asada}}, \citenamefont {{Azulay}},
  \citenamefont {{Baczko}}, \citenamefont {{Ball}}, \citenamefont
  {{Balokovi{\'c}}}, \citenamefont {{Barrett}},\ and\ \citenamefont
  {et~al.}}]{eht19-1}%
  \BibitemOpen
  \bibfield  {author} {\bibinfo {author} {\bibnamefont {{Event Horizon
  Telescope Collaboration}}}, \bibinfo {author} {\bibfnamefont
  {K.}~\bibnamefont {{Akiyama}}}, \bibinfo {author} {\bibfnamefont
  {A.}~\bibnamefont {{Alberdi}}}, \bibinfo {author} {\bibfnamefont
  {W.}~\bibnamefont {{Alef}}}, \bibinfo {author} {\bibfnamefont
  {K.}~\bibnamefont {{Asada}}}, \bibinfo {author} {\bibfnamefont
  {R.}~\bibnamefont {{Azulay}}}, \bibinfo {author} {\bibfnamefont {A.-K.}\
  \bibnamefont {{Baczko}}}, \bibinfo {author} {\bibfnamefont {D.}~\bibnamefont
  {{Ball}}}, \bibinfo {author} {\bibfnamefont {M.}~\bibnamefont
  {{Balokovi{\'c}}}}, \bibinfo {author} {\bibfnamefont {J.}~\bibnamefont
  {{Barrett}}},\ and\ \bibinfo {author} {\bibnamefont {et~al.}},\ }\bibfield
  {title} {\bibinfo {title} {{First M87 Event Horizon Telescope Results. I. The
  Shadow of the Supermassive Black Hole}},\ }\href
  {https://doi.org/10.3847/2041-8213/ab0ec7} {\bibfield  {journal} {\bibinfo
  {journal} {\apjl}\ }\textbf {\bibinfo {volume} {875}},\ \bibinfo {eid} {L1}
  (\bibinfo {year} {2019}{\natexlab{a}})},\ \Eprint
  {https://arxiv.org/abs/1906.11238} {arXiv:1906.11238 [astro-ph.GA]}
  \BibitemShut {NoStop}%
\bibitem [{\citenamefont {{Event Horizon Telescope Collaboration}}\ \emph
  {et~al.}(2019{\natexlab{b}})\citenamefont {{Event Horizon Telescope
  Collaboration}}, \citenamefont {{Akiyama}}, \citenamefont {{Alberdi}},
  \citenamefont {{Alef}}, \citenamefont {{Asada}}, \citenamefont {{Azulay}},
  \citenamefont {{Baczko}}, \citenamefont {{Ball}}, \citenamefont
  {{Balokovi{\'c}}}, \citenamefont {{Barrett}},\ and\ \citenamefont
  {et~al.}}]{eht19-2}%
  \BibitemOpen
  \bibfield  {author} {\bibinfo {author} {\bibnamefont {{Event Horizon
  Telescope Collaboration}}}, \bibinfo {author} {\bibfnamefont
  {K.}~\bibnamefont {{Akiyama}}}, \bibinfo {author} {\bibfnamefont
  {A.}~\bibnamefont {{Alberdi}}}, \bibinfo {author} {\bibfnamefont
  {W.}~\bibnamefont {{Alef}}}, \bibinfo {author} {\bibfnamefont
  {K.}~\bibnamefont {{Asada}}}, \bibinfo {author} {\bibfnamefont
  {R.}~\bibnamefont {{Azulay}}}, \bibinfo {author} {\bibfnamefont {A.-K.}\
  \bibnamefont {{Baczko}}}, \bibinfo {author} {\bibfnamefont {D.}~\bibnamefont
  {{Ball}}}, \bibinfo {author} {\bibfnamefont {M.}~\bibnamefont
  {{Balokovi{\'c}}}}, \bibinfo {author} {\bibfnamefont {J.}~\bibnamefont
  {{Barrett}}},\ and\ \bibinfo {author} {\bibnamefont {et~al.}},\ }\bibfield
  {title} {\bibinfo {title} {{First M87 Event Horizon Telescope Results. II.
  Array and Instrumentation}},\ }\href
  {https://doi.org/10.3847/2041-8213/ab0c96} {\bibfield  {journal} {\bibinfo
  {journal} {\apjl}\ }\textbf {\bibinfo {volume} {875}},\ \bibinfo {eid} {L2}
  (\bibinfo {year} {2019}{\natexlab{b}})},\ \Eprint
  {https://arxiv.org/abs/1906.11239} {arXiv:1906.11239 [astro-ph.IM]}
  \BibitemShut {NoStop}%
\bibitem [{\citenamefont {{Event Horizon Telescope Collaboration}}\ \emph
  {et~al.}(2019{\natexlab{c}})\citenamefont {{Event Horizon Telescope
  Collaboration}}, \citenamefont {{Akiyama}}, \citenamefont {{Alberdi}},
  \citenamefont {{Alef}}, \citenamefont {{Asada}}, \citenamefont {{Azulay}},
  \citenamefont {{Baczko}}, \citenamefont {{Ball}}, \citenamefont
  {{Balokovi{\'c}}}, \citenamefont {{Barrett}},\ and\ \citenamefont
  {et~al.}}]{eht19-3}%
  \BibitemOpen
  \bibfield  {author} {\bibinfo {author} {\bibnamefont {{Event Horizon
  Telescope Collaboration}}}, \bibinfo {author} {\bibfnamefont
  {K.}~\bibnamefont {{Akiyama}}}, \bibinfo {author} {\bibfnamefont
  {A.}~\bibnamefont {{Alberdi}}}, \bibinfo {author} {\bibfnamefont
  {W.}~\bibnamefont {{Alef}}}, \bibinfo {author} {\bibfnamefont
  {K.}~\bibnamefont {{Asada}}}, \bibinfo {author} {\bibfnamefont
  {R.}~\bibnamefont {{Azulay}}}, \bibinfo {author} {\bibfnamefont {A.-K.}\
  \bibnamefont {{Baczko}}}, \bibinfo {author} {\bibfnamefont {D.}~\bibnamefont
  {{Ball}}}, \bibinfo {author} {\bibfnamefont {M.}~\bibnamefont
  {{Balokovi{\'c}}}}, \bibinfo {author} {\bibfnamefont {J.}~\bibnamefont
  {{Barrett}}},\ and\ \bibinfo {author} {\bibnamefont {et~al.}},\ }\bibfield
  {title} {\bibinfo {title} {{First M87 Event Horizon Telescope Results. III.
  Data Processing and Calibration}},\ }\href
  {https://doi.org/10.3847/2041-8213/ab0c57} {\bibfield  {journal} {\bibinfo
  {journal} {\apjl}\ }\textbf {\bibinfo {volume} {875}},\ \bibinfo {eid} {L3}
  (\bibinfo {year} {2019}{\natexlab{c}})},\ \Eprint
  {https://arxiv.org/abs/1906.11240} {arXiv:1906.11240 [astro-ph.GA]}
  \BibitemShut {NoStop}%
\bibitem [{\citenamefont {{Event Horizon Telescope Collaboration}}\ \emph
  {et~al.}(2019{\natexlab{d}})\citenamefont {{Event Horizon Telescope
  Collaboration}}, \citenamefont {{Akiyama}}, \citenamefont {{Alberdi}},
  \citenamefont {{Alef}}, \citenamefont {{Asada}}, \citenamefont {{Azulay}},
  \citenamefont {{Baczko}}, \citenamefont {{Ball}}, \citenamefont
  {{Balokovi{\'c}}}, \citenamefont {{Barrett}},\ and\ \citenamefont
  {et~al.}}]{eht19-4}%
  \BibitemOpen
  \bibfield  {author} {\bibinfo {author} {\bibnamefont {{Event Horizon
  Telescope Collaboration}}}, \bibinfo {author} {\bibfnamefont
  {K.}~\bibnamefont {{Akiyama}}}, \bibinfo {author} {\bibfnamefont
  {A.}~\bibnamefont {{Alberdi}}}, \bibinfo {author} {\bibfnamefont
  {W.}~\bibnamefont {{Alef}}}, \bibinfo {author} {\bibfnamefont
  {K.}~\bibnamefont {{Asada}}}, \bibinfo {author} {\bibfnamefont
  {R.}~\bibnamefont {{Azulay}}}, \bibinfo {author} {\bibfnamefont {A.-K.}\
  \bibnamefont {{Baczko}}}, \bibinfo {author} {\bibfnamefont {D.}~\bibnamefont
  {{Ball}}}, \bibinfo {author} {\bibfnamefont {M.}~\bibnamefont
  {{Balokovi{\'c}}}}, \bibinfo {author} {\bibfnamefont {J.}~\bibnamefont
  {{Barrett}}},\ and\ \bibinfo {author} {\bibnamefont {et~al.}},\ }\bibfield
  {title} {\bibinfo {title} {{First M87 Event Horizon Telescope Results. IV.
  Imaging the Central Supermassive Black Hole}},\ }\href
  {https://doi.org/10.3847/2041-8213/ab0e85} {\bibfield  {journal} {\bibinfo
  {journal} {\apjl}\ }\textbf {\bibinfo {volume} {875}},\ \bibinfo {eid} {L4}
  (\bibinfo {year} {2019}{\natexlab{d}})},\ \Eprint
  {https://arxiv.org/abs/1906.11241} {arXiv:1906.11241 [astro-ph.GA]}
  \BibitemShut {NoStop}%
\bibitem [{\citenamefont {{Event Horizon Telescope Collaboration}}\ \emph
  {et~al.}(2019{\natexlab{e}})\citenamefont {{Event Horizon Telescope
  Collaboration}}, \citenamefont {{Akiyama}}, \citenamefont {{Alberdi}},
  \citenamefont {{Alef}}, \citenamefont {{Asada}}, \citenamefont {{Azulay}},
  \citenamefont {{Baczko}}, \citenamefont {{Ball}}, \citenamefont
  {{Balokovi{\'c}}}, \citenamefont {{Barrett}},\ and\ \citenamefont
  {et~al.}}]{eht19-5}%
  \BibitemOpen
  \bibfield  {author} {\bibinfo {author} {\bibnamefont {{Event Horizon
  Telescope Collaboration}}}, \bibinfo {author} {\bibfnamefont
  {K.}~\bibnamefont {{Akiyama}}}, \bibinfo {author} {\bibfnamefont
  {A.}~\bibnamefont {{Alberdi}}}, \bibinfo {author} {\bibfnamefont
  {W.}~\bibnamefont {{Alef}}}, \bibinfo {author} {\bibfnamefont
  {K.}~\bibnamefont {{Asada}}}, \bibinfo {author} {\bibfnamefont
  {R.}~\bibnamefont {{Azulay}}}, \bibinfo {author} {\bibfnamefont {A.-K.}\
  \bibnamefont {{Baczko}}}, \bibinfo {author} {\bibfnamefont {D.}~\bibnamefont
  {{Ball}}}, \bibinfo {author} {\bibfnamefont {M.}~\bibnamefont
  {{Balokovi{\'c}}}}, \bibinfo {author} {\bibfnamefont {J.}~\bibnamefont
  {{Barrett}}},\ and\ \bibinfo {author} {\bibnamefont {et~al.}},\ }\bibfield
  {title} {\bibinfo {title} {{First M87 Event Horizon Telescope Results. V.
  Physical Origin of the Asymmetric Ring}},\ }\href
  {https://doi.org/10.3847/2041-8213/ab0f43} {\bibfield  {journal} {\bibinfo
  {journal} {\apjl}\ }\textbf {\bibinfo {volume} {875}},\ \bibinfo {eid} {L5}
  (\bibinfo {year} {2019}{\natexlab{e}})},\ \Eprint
  {https://arxiv.org/abs/1906.11242} {arXiv:1906.11242 [astro-ph.GA]}
  \BibitemShut {NoStop}%
\bibitem [{\citenamefont {{Event Horizon Telescope Collaboration}}\ \emph
  {et~al.}(2019{\natexlab{f}})\citenamefont {{Event Horizon Telescope
  Collaboration}}, \citenamefont {{Akiyama}}, \citenamefont {{Alberdi}},
  \citenamefont {{Alef}}, \citenamefont {{Asada}}, \citenamefont {{Azulay}},
  \citenamefont {{Baczko}}, \citenamefont {{Ball}}, \citenamefont
  {{Balokovi{\'c}}}, \citenamefont {{Barrett}},\ and\ \citenamefont
  {et~al.}}]{eht19-6}%
  \BibitemOpen
  \bibfield  {author} {\bibinfo {author} {\bibnamefont {{Event Horizon
  Telescope Collaboration}}}, \bibinfo {author} {\bibfnamefont
  {K.}~\bibnamefont {{Akiyama}}}, \bibinfo {author} {\bibfnamefont
  {A.}~\bibnamefont {{Alberdi}}}, \bibinfo {author} {\bibfnamefont
  {W.}~\bibnamefont {{Alef}}}, \bibinfo {author} {\bibfnamefont
  {K.}~\bibnamefont {{Asada}}}, \bibinfo {author} {\bibfnamefont
  {R.}~\bibnamefont {{Azulay}}}, \bibinfo {author} {\bibfnamefont {A.-K.}\
  \bibnamefont {{Baczko}}}, \bibinfo {author} {\bibfnamefont {D.}~\bibnamefont
  {{Ball}}}, \bibinfo {author} {\bibfnamefont {M.}~\bibnamefont
  {{Balokovi{\'c}}}}, \bibinfo {author} {\bibfnamefont {J.}~\bibnamefont
  {{Barrett}}},\ and\ \bibinfo {author} {\bibnamefont {et~al.}},\ }\bibfield
  {title} {\bibinfo {title} {{First M87 Event Horizon Telescope Results. VI.
  The Shadow and Mass of the Central Black Hole}},\ }\href
  {https://doi.org/10.3847/2041-8213/ab1141} {\bibfield  {journal} {\bibinfo
  {journal} {\apjl}\ }\textbf {\bibinfo {volume} {875}},\ \bibinfo {eid} {L6}
  (\bibinfo {year} {2019}{\natexlab{f}})},\ \Eprint
  {https://arxiv.org/abs/1906.11243} {arXiv:1906.11243 [astro-ph.GA]}
  \BibitemShut {NoStop}%
\bibitem [{\citenamefont {{Event Horizon Telescope Collaboration}}\ \emph
  {et~al.}(2021)\citenamefont {{Event Horizon Telescope Collaboration}},
  \citenamefont {{Akiyama}}, \citenamefont {{Algaba}}, \citenamefont
  {{Alberdi}}, \citenamefont {{Alef}}, \citenamefont {{Anantua}}, \citenamefont
  {{Asada}}, \citenamefont {{Azulay}}, \citenamefont {{Baczko}}, \citenamefont
  {{Ball}},\ and\ \citenamefont {et~al.}}]{eht21-7}%
  \BibitemOpen
  \bibfield  {author} {\bibinfo {author} {\bibnamefont {{Event Horizon
  Telescope Collaboration}}}, \bibinfo {author} {\bibfnamefont
  {K.}~\bibnamefont {{Akiyama}}}, \bibinfo {author} {\bibfnamefont {J.~C.}\
  \bibnamefont {{Algaba}}}, \bibinfo {author} {\bibfnamefont {A.}~\bibnamefont
  {{Alberdi}}}, \bibinfo {author} {\bibfnamefont {W.}~\bibnamefont {{Alef}}},
  \bibinfo {author} {\bibfnamefont {R.}~\bibnamefont {{Anantua}}}, \bibinfo
  {author} {\bibfnamefont {K.}~\bibnamefont {{Asada}}}, \bibinfo {author}
  {\bibfnamefont {R.}~\bibnamefont {{Azulay}}}, \bibinfo {author}
  {\bibfnamefont {A.-K.}\ \bibnamefont {{Baczko}}}, \bibinfo {author}
  {\bibfnamefont {D.}~\bibnamefont {{Ball}}},\ and\ \bibinfo {author}
  {\bibnamefont {et~al.}},\ }\bibfield  {title} {\bibinfo {title} {{First M87
  Event Horizon Telescope Results. VII. Polarization of the Ring}},\ }\href
  {https://doi.org/10.3847/2041-8213/abe71d} {\bibfield  {journal} {\bibinfo
  {journal} {\apjl}\ }\textbf {\bibinfo {volume} {910}},\ \bibinfo {eid} {L12}
  (\bibinfo {year} {2021})},\ \Eprint {https://arxiv.org/abs/2105.01169}
  {arXiv:2105.01169 [astro-ph.HE]} \BibitemShut {NoStop}%
\bibitem [{\citenamefont {Janssen}\ \emph {et~al.}(2021)\citenamefont
  {Janssen}, \citenamefont {Falcke}, \citenamefont {Kadler}, \citenamefont
  {Ros}, \citenamefont {Wielgus}, \citenamefont {Akiyama}, \citenamefont
  {Balokovi{\'{c}}}, \citenamefont {Blackburn}, \citenamefont {Bouman},
  \citenamefont {Chael}, \citenamefont {Chan}, \citenamefont {Chatterjee},
  \citenamefont {Davelaar}, \citenamefont {Edwards}, \citenamefont {Fromm},
  \citenamefont {G{\'o}mez}, \citenamefont {Goddi}, \citenamefont {Issaoun},
  \citenamefont {Johnson}, \citenamefont {Kim}, \citenamefont {Koay},
  \citenamefont {Krichbaum}, \citenamefont {Liu}, \citenamefont {Liuzzo},
  \citenamefont {Markoff}, \citenamefont {Markowitz}, \citenamefont {Marrone},
  \citenamefont {Mizuno}, \citenamefont {M{\"u}ller}, \citenamefont {Ni},
  \citenamefont {Pesce}, \citenamefont {Ramakrishnan}, \citenamefont {Roelofs},
  \citenamefont {Rygl}, \citenamefont {van Bemmel}, \citenamefont {Alberdi},
  \citenamefont {Alef}, \citenamefont {Algaba}, \citenamefont {Anantua},
  \citenamefont {Asada}, \citenamefont {Azulay}, \citenamefont {Baczko},
  \citenamefont {Ball}, \citenamefont {Barrett}, \citenamefont {Benson},
  \citenamefont {Bintley}, \citenamefont {Blundell}, \citenamefont {Boland},
  \citenamefont {Bower}, \citenamefont {Boyce}, \citenamefont {Bremer},
  \citenamefont {Brinkerink}, \citenamefont {Brissenden}, \citenamefont
  {Britzen}, \citenamefont {Broderick}, \citenamefont {Broguiere},
  \citenamefont {Bronzwaer}, \citenamefont {Byun}, \citenamefont {Carlstrom},
  \citenamefont {Chatterjee}, \citenamefont {Chen}, \citenamefont {Chen},
  \citenamefont {Chesler}, \citenamefont {Cho}, \citenamefont {Christian},
  \citenamefont {Conway}, \citenamefont {Cordes}, \citenamefont {Crawford},
  \citenamefont {Crew}, \citenamefont {Cruz-Osorio}, \citenamefont {Cui},
  \citenamefont {De~Laurentis}, \citenamefont {Deane}, \citenamefont {Dempsey},
  \citenamefont {Desvignes}, \citenamefont {Dexter}, \citenamefont {Doeleman},
  \citenamefont {Eatough}, \citenamefont {Farah}, \citenamefont {Fish},
  \citenamefont {Fomalont}, \citenamefont {Ford}, \citenamefont
  {Fraga-Encinas}, \citenamefont {Friberg}, \citenamefont {Fuentes},
  \citenamefont {Galison}, \citenamefont {Gammie}, \citenamefont {Garc{\'i}a},
  \citenamefont {Gelles}, \citenamefont {Gentaz}, \citenamefont {Georgiev},
  \citenamefont {Gold}, \citenamefont {G{\'o}mez-Ruiz}, \citenamefont {Gu},
  \citenamefont {Gurwell}, \citenamefont {Hada}, \citenamefont {Haggard},
  \citenamefont {Hecht}, \citenamefont {Hesper}, \citenamefont {Himwich},
  \citenamefont {Ho}, \citenamefont {Ho}, \citenamefont {Honma}, \citenamefont
  {Huang}, \citenamefont {Huang}, \citenamefont {Hughes}, \citenamefont
  {Ikeda}, \citenamefont {Inoue}, \citenamefont {James}, \citenamefont
  {Jannuzi}, \citenamefont {Jeter}, \citenamefont {Jiang}, \citenamefont
  {Jimenez-Rosales}, \citenamefont {Jorstad}, \citenamefont {Jung},
  \citenamefont {Karami}, \citenamefont {Karuppusamy}, \citenamefont
  {Kawashima}, \citenamefont {Keating}, \citenamefont {Kettenis}, \citenamefont
  {Kim}, \citenamefont {Kim}, \citenamefont {Kim}, \citenamefont {Kino},
  \citenamefont {Kofuji}, \citenamefont {Koyama}, \citenamefont {Kramer},
  \citenamefont {Kramer}, \citenamefont {Kuo}, \citenamefont {Lauer},
  \citenamefont {Lee}, \citenamefont {Levis}, \citenamefont {Li}, \citenamefont
  {Li}, \citenamefont {Lindqvist}, \citenamefont {Lico}, \citenamefont
  {Lindahl}, \citenamefont {Liu}, \citenamefont {Lo}, \citenamefont {Lobanov},
  \citenamefont {Loinard}, \citenamefont {Lonsdale}, \citenamefont {Lu},
  \citenamefont {MacDonald}, \citenamefont {Mao}, \citenamefont {Marchili},
  \citenamefont {Marscher}, \citenamefont {Mart{\'i}-Vidal}, \citenamefont
  {Matsushita}, \citenamefont {Matthews}, \citenamefont {Medeiros},
  \citenamefont {Menten}, \citenamefont {Mizuno}, \citenamefont {Moran},
  \citenamefont {Moriyama}, \citenamefont {Moscibrodzka}, \citenamefont
  {Musoke}, \citenamefont {Mej{\'i}as}, \citenamefont {Nagai}, \citenamefont
  {Nagar}, \citenamefont {Nakamura}, \citenamefont {Narayan}, \citenamefont
  {Narayanan}, \citenamefont {Natarajan}, \citenamefont {Nathanail},
  \citenamefont {Neilsen}, \citenamefont {Neri}, \citenamefont {Noutsos},
  \citenamefont {Nowak}, \citenamefont {Okino}, \citenamefont {Olivares},
  \citenamefont {Ortiz-Le{\'o}n}, \citenamefont {Oyama}, \citenamefont
  {{\"O}zel}, \citenamefont {Palumbo}, \citenamefont {Park}, \citenamefont
  {Patel}, \citenamefont {Pen}, \citenamefont {Pi{\'e}tu}, \citenamefont
  {Plambeck}, \citenamefont {PopStefanija}, \citenamefont {Porth},
  \citenamefont {P{\"o}tzl}, \citenamefont {Prather}, \citenamefont
  {Preciado-L{\'o}pez}, \citenamefont {Psaltis}, \citenamefont {Pu},
  \citenamefont {Rao}, \citenamefont {Rawlings}, \citenamefont {Raymond},
  \citenamefont {Rezzolla}, \citenamefont {Ricarte}, \citenamefont {Ripperda},
  \citenamefont {Rogers}, \citenamefont {Rose}, \citenamefont {Roshanineshat},
  \citenamefont {Rottmann}, \citenamefont {Roy}, \citenamefont {Ruszczyk},
  \citenamefont {S{\'a}nchez}, \citenamefont {S{\'a}nchez-Arguelles},
  \citenamefont {Sasada}, \citenamefont {Savolainen}, \citenamefont {Schloerb},
  \citenamefont {Schuster}, \citenamefont {Shao}, \citenamefont {Shen},
  \citenamefont {Small}, \citenamefont {Sohn}, \citenamefont {SooHoo},
  \citenamefont {Sun}, \citenamefont {Tazaki}, \citenamefont {Tetarenko},
  \citenamefont {Tiede}, \citenamefont {Tilanus}, \citenamefont {Titus},
  \citenamefont {Torne}, \citenamefont {Trent}, \citenamefont {Traianou},
  \citenamefont {Trippe}, \citenamefont {van Langevelde}, \citenamefont {van
  Rossum}, \citenamefont {Wagner}, \citenamefont {Ward-Thompson}, \citenamefont
  {Wardle}, \citenamefont {Weintroub}, \citenamefont {Wex}, \citenamefont
  {Wharton}, \citenamefont {Wong}, \citenamefont {Wu}, \citenamefont {Yoon},
  \citenamefont {Young}, \citenamefont {Young}, \citenamefont {Younsi},
  \citenamefont {Yuan}, \citenamefont {Yuan}, \citenamefont {Zensus},
  \citenamefont {Zhao}, \citenamefont {Zhao},\ and\ \citenamefont
  {Collaboration}}]{Janssen+2021}%
  \BibitemOpen
  \bibfield  {author} {\bibinfo {author} {\bibfnamefont {M.}~\bibnamefont
  {Janssen}}, \bibinfo {author} {\bibfnamefont {H.}~\bibnamefont {Falcke}},
  \bibinfo {author} {\bibfnamefont {M.}~\bibnamefont {Kadler}}, \bibinfo
  {author} {\bibfnamefont {E.}~\bibnamefont {Ros}}, \bibinfo {author}
  {\bibfnamefont {M.}~\bibnamefont {Wielgus}}, \bibinfo {author} {\bibfnamefont
  {K.}~\bibnamefont {Akiyama}}, \bibinfo {author} {\bibfnamefont
  {M.}~\bibnamefont {Balokovi{\'{c}}}}, \bibinfo {author} {\bibfnamefont
  {L.}~\bibnamefont {Blackburn}}, \bibinfo {author} {\bibfnamefont {K.~L.}\
  \bibnamefont {Bouman}}, \bibinfo {author} {\bibfnamefont {A.}~\bibnamefont
  {Chael}}, \bibinfo {author} {\bibfnamefont {C.-k.}\ \bibnamefont {Chan}},
  \bibinfo {author} {\bibfnamefont {K.}~\bibnamefont {Chatterjee}}, \bibinfo
  {author} {\bibfnamefont {J.}~\bibnamefont {Davelaar}}, \bibinfo {author}
  {\bibfnamefont {P.~G.}\ \bibnamefont {Edwards}}, \bibinfo {author}
  {\bibfnamefont {C.~M.}\ \bibnamefont {Fromm}}, \bibinfo {author}
  {\bibfnamefont {J.~L.}\ \bibnamefont {G{\'o}mez}}, \bibinfo {author}
  {\bibfnamefont {C.}~\bibnamefont {Goddi}}, \bibinfo {author} {\bibfnamefont
  {S.}~\bibnamefont {Issaoun}}, \bibinfo {author} {\bibfnamefont {M.~D.}\
  \bibnamefont {Johnson}}, \bibinfo {author} {\bibfnamefont {J.}~\bibnamefont
  {Kim}}, \bibinfo {author} {\bibfnamefont {J.~Y.}\ \bibnamefont {Koay}},
  \bibinfo {author} {\bibfnamefont {T.~P.}\ \bibnamefont {Krichbaum}}, \bibinfo
  {author} {\bibfnamefont {J.}~\bibnamefont {Liu}}, \bibinfo {author}
  {\bibfnamefont {E.}~\bibnamefont {Liuzzo}}, \bibinfo {author} {\bibfnamefont
  {S.}~\bibnamefont {Markoff}}, \bibinfo {author} {\bibfnamefont
  {A.}~\bibnamefont {Markowitz}}, \bibinfo {author} {\bibfnamefont {D.~P.}\
  \bibnamefont {Marrone}}, \bibinfo {author} {\bibfnamefont {Y.}~\bibnamefont
  {Mizuno}}, \bibinfo {author} {\bibfnamefont {C.}~\bibnamefont {M{\"u}ller}},
  \bibinfo {author} {\bibfnamefont {C.}~\bibnamefont {Ni}}, \bibinfo {author}
  {\bibfnamefont {D.~W.}\ \bibnamefont {Pesce}}, \bibinfo {author}
  {\bibfnamefont {V.}~\bibnamefont {Ramakrishnan}}, \bibinfo {author}
  {\bibfnamefont {F.}~\bibnamefont {Roelofs}}, \bibinfo {author} {\bibfnamefont
  {K.~L.~J.}\ \bibnamefont {Rygl}}, \bibinfo {author} {\bibfnamefont
  {I.}~\bibnamefont {van Bemmel}}, \bibinfo {author} {\bibfnamefont
  {A.}~\bibnamefont {Alberdi}}, \bibinfo {author} {\bibfnamefont
  {W.}~\bibnamefont {Alef}}, \bibinfo {author} {\bibfnamefont {J.~C.}\
  \bibnamefont {Algaba}}, \bibinfo {author} {\bibfnamefont {R.}~\bibnamefont
  {Anantua}}, \bibinfo {author} {\bibfnamefont {K.}~\bibnamefont {Asada}},
  \bibinfo {author} {\bibfnamefont {R.}~\bibnamefont {Azulay}}, \bibinfo
  {author} {\bibfnamefont {A.-K.}\ \bibnamefont {Baczko}}, \bibinfo {author}
  {\bibfnamefont {D.}~\bibnamefont {Ball}}, \bibinfo {author} {\bibfnamefont
  {J.}~\bibnamefont {Barrett}}, \bibinfo {author} {\bibfnamefont {B.~A.}\
  \bibnamefont {Benson}}, \bibinfo {author} {\bibfnamefont {D.}~\bibnamefont
  {Bintley}}, \bibinfo {author} {\bibfnamefont {R.}~\bibnamefont {Blundell}},
  \bibinfo {author} {\bibfnamefont {W.}~\bibnamefont {Boland}}, \bibinfo
  {author} {\bibfnamefont {G.~C.}\ \bibnamefont {Bower}}, \bibinfo {author}
  {\bibfnamefont {H.}~\bibnamefont {Boyce}}, \bibinfo {author} {\bibfnamefont
  {M.}~\bibnamefont {Bremer}}, \bibinfo {author} {\bibfnamefont {C.~D.}\
  \bibnamefont {Brinkerink}}, \bibinfo {author} {\bibfnamefont
  {R.}~\bibnamefont {Brissenden}}, \bibinfo {author} {\bibfnamefont
  {S.}~\bibnamefont {Britzen}}, \bibinfo {author} {\bibfnamefont {A.~E.}\
  \bibnamefont {Broderick}}, \bibinfo {author} {\bibfnamefont {D.}~\bibnamefont
  {Broguiere}}, \bibinfo {author} {\bibfnamefont {T.}~\bibnamefont
  {Bronzwaer}}, \bibinfo {author} {\bibfnamefont {D.-Y.}\ \bibnamefont {Byun}},
  \bibinfo {author} {\bibfnamefont {J.~E.}\ \bibnamefont {Carlstrom}}, \bibinfo
  {author} {\bibfnamefont {S.}~\bibnamefont {Chatterjee}}, \bibinfo {author}
  {\bibfnamefont {M.-T.}\ \bibnamefont {Chen}}, \bibinfo {author}
  {\bibfnamefont {Y.}~\bibnamefont {Chen}}, \bibinfo {author} {\bibfnamefont
  {P.~M.}\ \bibnamefont {Chesler}}, \bibinfo {author} {\bibfnamefont
  {I.}~\bibnamefont {Cho}}, \bibinfo {author} {\bibfnamefont {P.}~\bibnamefont
  {Christian}}, \bibinfo {author} {\bibfnamefont {J.~E.}\ \bibnamefont
  {Conway}}, \bibinfo {author} {\bibfnamefont {J.~M.}\ \bibnamefont {Cordes}},
  \bibinfo {author} {\bibfnamefont {T.~M.}\ \bibnamefont {Crawford}}, \bibinfo
  {author} {\bibfnamefont {G.~B.}\ \bibnamefont {Crew}}, \bibinfo {author}
  {\bibfnamefont {A.}~\bibnamefont {Cruz-Osorio}}, \bibinfo {author}
  {\bibfnamefont {Y.}~\bibnamefont {Cui}}, \bibinfo {author} {\bibfnamefont
  {M.}~\bibnamefont {De~Laurentis}}, \bibinfo {author} {\bibfnamefont
  {R.}~\bibnamefont {Deane}}, \bibinfo {author} {\bibfnamefont
  {J.}~\bibnamefont {Dempsey}}, \bibinfo {author} {\bibfnamefont
  {G.}~\bibnamefont {Desvignes}}, \bibinfo {author} {\bibfnamefont
  {J.}~\bibnamefont {Dexter}}, \bibinfo {author} {\bibfnamefont {S.~S.}\
  \bibnamefont {Doeleman}}, \bibinfo {author} {\bibfnamefont {R.~P.}\
  \bibnamefont {Eatough}}, \bibinfo {author} {\bibfnamefont {J.}~\bibnamefont
  {Farah}}, \bibinfo {author} {\bibfnamefont {V.~L.}\ \bibnamefont {Fish}},
  \bibinfo {author} {\bibfnamefont {E.}~\bibnamefont {Fomalont}}, \bibinfo
  {author} {\bibfnamefont {H.~A.}\ \bibnamefont {Ford}}, \bibinfo {author}
  {\bibfnamefont {R.}~\bibnamefont {Fraga-Encinas}}, \bibinfo {author}
  {\bibfnamefont {P.}~\bibnamefont {Friberg}}, \bibinfo {author} {\bibfnamefont
  {A.}~\bibnamefont {Fuentes}}, \bibinfo {author} {\bibfnamefont
  {P.}~\bibnamefont {Galison}}, \bibinfo {author} {\bibfnamefont {C.~F.}\
  \bibnamefont {Gammie}}, \bibinfo {author} {\bibfnamefont {R.}~\bibnamefont
  {Garc{\'i}a}}, \bibinfo {author} {\bibfnamefont {Z.}~\bibnamefont {Gelles}},
  \bibinfo {author} {\bibfnamefont {O.}~\bibnamefont {Gentaz}}, \bibinfo
  {author} {\bibfnamefont {B.}~\bibnamefont {Georgiev}}, \bibinfo {author}
  {\bibfnamefont {R.}~\bibnamefont {Gold}}, \bibinfo {author} {\bibfnamefont
  {A.~I.}\ \bibnamefont {G{\'o}mez-Ruiz}}, \bibinfo {author} {\bibfnamefont
  {M.}~\bibnamefont {Gu}}, \bibinfo {author} {\bibfnamefont {M.}~\bibnamefont
  {Gurwell}}, \bibinfo {author} {\bibfnamefont {K.}~\bibnamefont {Hada}},
  \bibinfo {author} {\bibfnamefont {D.}~\bibnamefont {Haggard}}, \bibinfo
  {author} {\bibfnamefont {M.~H.}\ \bibnamefont {Hecht}}, \bibinfo {author}
  {\bibfnamefont {R.}~\bibnamefont {Hesper}}, \bibinfo {author} {\bibfnamefont
  {E.}~\bibnamefont {Himwich}}, \bibinfo {author} {\bibfnamefont {L.~C.}\
  \bibnamefont {Ho}}, \bibinfo {author} {\bibfnamefont {P.}~\bibnamefont {Ho}},
  \bibinfo {author} {\bibfnamefont {M.}~\bibnamefont {Honma}}, \bibinfo
  {author} {\bibfnamefont {C.-W.~L.}\ \bibnamefont {Huang}}, \bibinfo {author}
  {\bibfnamefont {L.}~\bibnamefont {Huang}}, \bibinfo {author} {\bibfnamefont
  {D.~H.}\ \bibnamefont {Hughes}}, \bibinfo {author} {\bibfnamefont
  {S.}~\bibnamefont {Ikeda}}, \bibinfo {author} {\bibfnamefont
  {M.}~\bibnamefont {Inoue}}, \bibinfo {author} {\bibfnamefont {D.~J.}\
  \bibnamefont {James}}, \bibinfo {author} {\bibfnamefont {B.~T.}\ \bibnamefont
  {Jannuzi}}, \bibinfo {author} {\bibfnamefont {B.}~\bibnamefont {Jeter}},
  \bibinfo {author} {\bibfnamefont {W.}~\bibnamefont {Jiang}}, \bibinfo
  {author} {\bibfnamefont {A.}~\bibnamefont {Jimenez-Rosales}}, \bibinfo
  {author} {\bibfnamefont {S.}~\bibnamefont {Jorstad}}, \bibinfo {author}
  {\bibfnamefont {T.}~\bibnamefont {Jung}}, \bibinfo {author} {\bibfnamefont
  {M.}~\bibnamefont {Karami}}, \bibinfo {author} {\bibfnamefont
  {R.}~\bibnamefont {Karuppusamy}}, \bibinfo {author} {\bibfnamefont
  {T.}~\bibnamefont {Kawashima}}, \bibinfo {author} {\bibfnamefont {G.~K.}\
  \bibnamefont {Keating}}, \bibinfo {author} {\bibfnamefont {M.}~\bibnamefont
  {Kettenis}}, \bibinfo {author} {\bibfnamefont {D.-J.}\ \bibnamefont {Kim}},
  \bibinfo {author} {\bibfnamefont {J.-Y.}\ \bibnamefont {Kim}}, \bibinfo
  {author} {\bibfnamefont {J.}~\bibnamefont {Kim}}, \bibinfo {author}
  {\bibfnamefont {M.}~\bibnamefont {Kino}}, \bibinfo {author} {\bibfnamefont
  {Y.}~\bibnamefont {Kofuji}}, \bibinfo {author} {\bibfnamefont
  {S.}~\bibnamefont {Koyama}}, \bibinfo {author} {\bibfnamefont
  {M.}~\bibnamefont {Kramer}}, \bibinfo {author} {\bibfnamefont
  {C.}~\bibnamefont {Kramer}}, \bibinfo {author} {\bibfnamefont {C.-Y.}\
  \bibnamefont {Kuo}}, \bibinfo {author} {\bibfnamefont {T.~R.}\ \bibnamefont
  {Lauer}}, \bibinfo {author} {\bibfnamefont {S.-S.}\ \bibnamefont {Lee}},
  \bibinfo {author} {\bibfnamefont {A.}~\bibnamefont {Levis}}, \bibinfo
  {author} {\bibfnamefont {Y.-R.}\ \bibnamefont {Li}}, \bibinfo {author}
  {\bibfnamefont {Z.}~\bibnamefont {Li}}, \bibinfo {author} {\bibfnamefont
  {M.}~\bibnamefont {Lindqvist}}, \bibinfo {author} {\bibfnamefont
  {R.}~\bibnamefont {Lico}}, \bibinfo {author} {\bibfnamefont {G.}~\bibnamefont
  {Lindahl}}, \bibinfo {author} {\bibfnamefont {K.}~\bibnamefont {Liu}},
  \bibinfo {author} {\bibfnamefont {W.-P.}\ \bibnamefont {Lo}}, \bibinfo
  {author} {\bibfnamefont {A.~P.}\ \bibnamefont {Lobanov}}, \bibinfo {author}
  {\bibfnamefont {L.}~\bibnamefont {Loinard}}, \bibinfo {author} {\bibfnamefont
  {C.}~\bibnamefont {Lonsdale}}, \bibinfo {author} {\bibfnamefont {R.-S.}\
  \bibnamefont {Lu}}, \bibinfo {author} {\bibfnamefont {N.~R.}\ \bibnamefont
  {MacDonald}}, \bibinfo {author} {\bibfnamefont {J.}~\bibnamefont {Mao}},
  \bibinfo {author} {\bibfnamefont {N.}~\bibnamefont {Marchili}}, \bibinfo
  {author} {\bibfnamefont {A.~P.}\ \bibnamefont {Marscher}}, \bibinfo {author}
  {\bibfnamefont {I.}~\bibnamefont {Mart{\'i}-Vidal}}, \bibinfo {author}
  {\bibfnamefont {S.}~\bibnamefont {Matsushita}}, \bibinfo {author}
  {\bibfnamefont {L.~D.}\ \bibnamefont {Matthews}}, \bibinfo {author}
  {\bibfnamefont {L.}~\bibnamefont {Medeiros}}, \bibinfo {author}
  {\bibfnamefont {K.~M.}\ \bibnamefont {Menten}}, \bibinfo {author}
  {\bibfnamefont {I.}~\bibnamefont {Mizuno}}, \bibinfo {author} {\bibfnamefont
  {J.~M.}\ \bibnamefont {Moran}}, \bibinfo {author} {\bibfnamefont
  {K.}~\bibnamefont {Moriyama}}, \bibinfo {author} {\bibfnamefont
  {M.}~\bibnamefont {Moscibrodzka}}, \bibinfo {author} {\bibfnamefont
  {G.}~\bibnamefont {Musoke}}, \bibinfo {author} {\bibfnamefont {A.~M.}\
  \bibnamefont {Mej{\'i}as}}, \bibinfo {author} {\bibfnamefont
  {H.}~\bibnamefont {Nagai}}, \bibinfo {author} {\bibfnamefont {N.~M.}\
  \bibnamefont {Nagar}}, \bibinfo {author} {\bibfnamefont {M.}~\bibnamefont
  {Nakamura}}, \bibinfo {author} {\bibfnamefont {R.}~\bibnamefont {Narayan}},
  \bibinfo {author} {\bibfnamefont {G.}~\bibnamefont {Narayanan}}, \bibinfo
  {author} {\bibfnamefont {I.}~\bibnamefont {Natarajan}}, \bibinfo {author}
  {\bibfnamefont {A.}~\bibnamefont {Nathanail}}, \bibinfo {author}
  {\bibfnamefont {J.}~\bibnamefont {Neilsen}}, \bibinfo {author} {\bibfnamefont
  {R.}~\bibnamefont {Neri}}, \bibinfo {author} {\bibfnamefont {A.}~\bibnamefont
  {Noutsos}}, \bibinfo {author} {\bibfnamefont {M.~A.}\ \bibnamefont {Nowak}},
  \bibinfo {author} {\bibfnamefont {H.}~\bibnamefont {Okino}}, \bibinfo
  {author} {\bibfnamefont {H.}~\bibnamefont {Olivares}}, \bibinfo {author}
  {\bibfnamefont {G.~N.}\ \bibnamefont {Ortiz-Le{\'o}n}}, \bibinfo {author}
  {\bibfnamefont {T.}~\bibnamefont {Oyama}}, \bibinfo {author} {\bibfnamefont
  {F.}~\bibnamefont {{\"O}zel}}, \bibinfo {author} {\bibfnamefont {D.~C.~M.}\
  \bibnamefont {Palumbo}}, \bibinfo {author} {\bibfnamefont {J.}~\bibnamefont
  {Park}}, \bibinfo {author} {\bibfnamefont {N.}~\bibnamefont {Patel}},
  \bibinfo {author} {\bibfnamefont {U.-L.}\ \bibnamefont {Pen}}, \bibinfo
  {author} {\bibfnamefont {V.}~\bibnamefont {Pi{\'e}tu}}, \bibinfo {author}
  {\bibfnamefont {R.}~\bibnamefont {Plambeck}}, \bibinfo {author}
  {\bibfnamefont {A.}~\bibnamefont {PopStefanija}}, \bibinfo {author}
  {\bibfnamefont {O.}~\bibnamefont {Porth}}, \bibinfo {author} {\bibfnamefont
  {F.~M.}\ \bibnamefont {P{\"o}tzl}}, \bibinfo {author} {\bibfnamefont
  {B.}~\bibnamefont {Prather}}, \bibinfo {author} {\bibfnamefont {J.~A.}\
  \bibnamefont {Preciado-L{\'o}pez}}, \bibinfo {author} {\bibfnamefont
  {D.}~\bibnamefont {Psaltis}}, \bibinfo {author} {\bibfnamefont {H.-Y.}\
  \bibnamefont {Pu}}, \bibinfo {author} {\bibfnamefont {R.}~\bibnamefont
  {Rao}}, \bibinfo {author} {\bibfnamefont {M.~G.}\ \bibnamefont {Rawlings}},
  \bibinfo {author} {\bibfnamefont {A.~W.}\ \bibnamefont {Raymond}}, \bibinfo
  {author} {\bibfnamefont {L.}~\bibnamefont {Rezzolla}}, \bibinfo {author}
  {\bibfnamefont {A.}~\bibnamefont {Ricarte}}, \bibinfo {author} {\bibfnamefont
  {B.}~\bibnamefont {Ripperda}}, \bibinfo {author} {\bibfnamefont
  {A.}~\bibnamefont {Rogers}}, \bibinfo {author} {\bibfnamefont
  {M.}~\bibnamefont {Rose}}, \bibinfo {author} {\bibfnamefont {A.}~\bibnamefont
  {Roshanineshat}}, \bibinfo {author} {\bibfnamefont {H.}~\bibnamefont
  {Rottmann}}, \bibinfo {author} {\bibfnamefont {A.~L.}\ \bibnamefont {Roy}},
  \bibinfo {author} {\bibfnamefont {C.}~\bibnamefont {Ruszczyk}}, \bibinfo
  {author} {\bibfnamefont {S.}~\bibnamefont {S{\'a}nchez}}, \bibinfo {author}
  {\bibfnamefont {D.}~\bibnamefont {S{\'a}nchez-Arguelles}}, \bibinfo {author}
  {\bibfnamefont {M.}~\bibnamefont {Sasada}}, \bibinfo {author} {\bibfnamefont
  {T.}~\bibnamefont {Savolainen}}, \bibinfo {author} {\bibfnamefont {F.~P.}\
  \bibnamefont {Schloerb}}, \bibinfo {author} {\bibfnamefont {K.-F.}\
  \bibnamefont {Schuster}}, \bibinfo {author} {\bibfnamefont {L.}~\bibnamefont
  {Shao}}, \bibinfo {author} {\bibfnamefont {Z.}~\bibnamefont {Shen}}, \bibinfo
  {author} {\bibfnamefont {D.}~\bibnamefont {Small}}, \bibinfo {author}
  {\bibfnamefont {B.~W.}\ \bibnamefont {Sohn}}, \bibinfo {author}
  {\bibfnamefont {J.}~\bibnamefont {SooHoo}}, \bibinfo {author} {\bibfnamefont
  {H.}~\bibnamefont {Sun}}, \bibinfo {author} {\bibfnamefont {F.}~\bibnamefont
  {Tazaki}}, \bibinfo {author} {\bibfnamefont {A.~J.}\ \bibnamefont
  {Tetarenko}}, \bibinfo {author} {\bibfnamefont {P.}~\bibnamefont {Tiede}},
  \bibinfo {author} {\bibfnamefont {R.~P.~J.}\ \bibnamefont {Tilanus}},
  \bibinfo {author} {\bibfnamefont {M.}~\bibnamefont {Titus}}, \bibinfo
  {author} {\bibfnamefont {P.}~\bibnamefont {Torne}}, \bibinfo {author}
  {\bibfnamefont {T.}~\bibnamefont {Trent}}, \bibinfo {author} {\bibfnamefont
  {E.}~\bibnamefont {Traianou}}, \bibinfo {author} {\bibfnamefont
  {S.}~\bibnamefont {Trippe}}, \bibinfo {author} {\bibfnamefont {H.~J.}\
  \bibnamefont {van Langevelde}}, \bibinfo {author} {\bibfnamefont {D.~R.}\
  \bibnamefont {van Rossum}}, \bibinfo {author} {\bibfnamefont
  {J.}~\bibnamefont {Wagner}}, \bibinfo {author} {\bibfnamefont
  {D.}~\bibnamefont {Ward-Thompson}}, \bibinfo {author} {\bibfnamefont
  {J.}~\bibnamefont {Wardle}}, \bibinfo {author} {\bibfnamefont
  {J.}~\bibnamefont {Weintroub}}, \bibinfo {author} {\bibfnamefont
  {N.}~\bibnamefont {Wex}}, \bibinfo {author} {\bibfnamefont {R.}~\bibnamefont
  {Wharton}}, \bibinfo {author} {\bibfnamefont {G.~N.}\ \bibnamefont {Wong}},
  \bibinfo {author} {\bibfnamefont {Q.}~\bibnamefont {Wu}}, \bibinfo {author}
  {\bibfnamefont {D.}~\bibnamefont {Yoon}}, \bibinfo {author} {\bibfnamefont
  {A.}~\bibnamefont {Young}}, \bibinfo {author} {\bibfnamefont
  {K.}~\bibnamefont {Young}}, \bibinfo {author} {\bibfnamefont
  {Z.}~\bibnamefont {Younsi}}, \bibinfo {author} {\bibfnamefont
  {F.}~\bibnamefont {Yuan}}, \bibinfo {author} {\bibfnamefont {Y.-F.}\
  \bibnamefont {Yuan}}, \bibinfo {author} {\bibfnamefont {J.~A.}\ \bibnamefont
  {Zensus}}, \bibinfo {author} {\bibfnamefont {G.-Y.}\ \bibnamefont {Zhao}},
  \bibinfo {author} {\bibfnamefont {S.-S.}\ \bibnamefont {Zhao}},\ and\
  \bibinfo {author} {\bibfnamefont {T.~E. H.~T.}\ \bibnamefont
  {Collaboration}},\ }\bibfield  {title} {\bibinfo {title} {Event horizon
  telescope observations of the jet launching and collimation in centaurus a},\
  }\bibfield  {journal} {\bibinfo  {journal} {Nature Astronomy}\ }\href
  {https://doi.org/10.1038/s41550-021-01417-w} {10.1038/s41550-021-01417-w}
  (\bibinfo {year} {2021})\BibitemShut {NoStop}%
\bibitem [{\citenamefont {Hauptman}(1986{\natexlab{a}})}]{Hauptman1986}%
  \BibitemOpen
  \bibfield  {author} {\bibinfo {author} {\bibfnamefont {H.}~\bibnamefont
  {Hauptman}},\ }\bibfield  {title} {\bibinfo {title} {The direct methods of
  x-ray crystallography},\ }\href
  {https://doi.org/10.1126/science.233.4760.178} {\bibfield  {journal}
  {\bibinfo  {journal} {Science}\ }\textbf {\bibinfo {volume} {233}},\ \bibinfo
  {pages} {178} (\bibinfo {year} {1986}{\natexlab{a}})},\ \Eprint
  {https://arxiv.org/abs/https://science.sciencemag.org/content/233/4760/178.full.pdf}
  {https://science.sciencemag.org/content/233/4760/178.full.pdf} \BibitemShut
  {NoStop}%
\bibitem [{\citenamefont {Hauptman}(1986{\natexlab{b}})}]{Hauptman-Nobel}%
  \BibitemOpen
  \bibfield  {author} {\bibinfo {author} {\bibfnamefont {H.}~\bibnamefont
  {Hauptman}},\ }\bibfield  {title} {\bibinfo {title} {Direct methods and
  anomalous dispersion (nobel lecture)},\ }\href
  {https://doi.org/10.1002/anie.198606031} {\bibfield  {journal} {\bibinfo
  {journal} {Angewandte Chemie International Edition in English}\ }\textbf
  {\bibinfo {volume} {25}},\ \bibinfo {pages} {603} (\bibinfo {year}
  {1986}{\natexlab{b}})},\ \Eprint
  {https://arxiv.org/abs/https://onlinelibrary.wiley.com/doi/pdf/10.1002/anie.198606031}
  {https://onlinelibrary.wiley.com/doi/pdf/10.1002/anie.198606031} \BibitemShut
  {NoStop}%
\bibitem [{\citenamefont {Hauptman}(1991)}]{Hauptman1991}%
  \BibitemOpen
  \bibfield  {author} {\bibinfo {author} {\bibfnamefont {H.~A.}\ \bibnamefont
  {Hauptman}},\ }\bibfield  {title} {\bibinfo {title} {The phase problem of
  x-ray crystallography},\ }\href {https://doi.org/10.1088/0034-4885/54/11/002}
  {\bibfield  {journal} {\bibinfo  {journal} {Reports on Progress in Physics}\
  }\textbf {\bibinfo {volume} {54}},\ \bibinfo {pages} {1427} (\bibinfo {year}
  {1991})}\BibitemShut {NoStop}%
\bibitem [{\citenamefont {Giacovazzo}\ \emph {et~al.}(2002)\citenamefont
  {Giacovazzo}, \citenamefont {Capitelli}, \citenamefont {Cuocci},\ and\
  \citenamefont {Ianigro}}]{Giacovazzo+2002}%
  \BibitemOpen
  \bibfield  {author} {\bibinfo {author} {\bibfnamefont {C.}~\bibnamefont
  {Giacovazzo}}, \bibinfo {author} {\bibfnamefont {F.}~\bibnamefont
  {Capitelli}}, \bibinfo {author} {\bibfnamefont {C.}~\bibnamefont {Cuocci}},\
  and\ \bibinfo {author} {\bibfnamefont {M.}~\bibnamefont {Ianigro}},\
  }\bibfield  {title} {\bibinfo {title} {Direct methods and applications to
  electron crystallography},\ }in\ \href
  {https://doi.org/https://doi.org/10.1016/S1076-5670(02)80067-2} {\emph
  {\bibinfo {booktitle} {Microscopy, Spectroscopy, Holography and
  Crystallography with Electrons}}},\ \bibinfo {series} {Advances in Imaging
  and Electron Physics}, Vol.\ \bibinfo {volume} {123},\ \bibinfo {editor}
  {edited by\ \bibinfo {editor} {\bibfnamefont {P.~W.}\ \bibnamefont {Hawkes}},
  \bibinfo {editor} {\bibfnamefont {P.~G.}\ \bibnamefont {Merli}}, \bibinfo
  {editor} {\bibfnamefont {G.}~\bibnamefont {Calestani}},\ and\ \bibinfo
  {editor} {\bibfnamefont {M.}~\bibnamefont {Vittori-Antisari}}}\ (\bibinfo
  {publisher} {Elsevier},\ \bibinfo {year} {2002})\ pp.\ \bibinfo {pages} {291
  -- 310}\BibitemShut {NoStop}%
\bibitem [{\citenamefont {Giacovazzo}(2014)}]{Giacovazzo2014}%
  \BibitemOpen
  \bibfield  {author} {\bibinfo {author} {\bibfnamefont {C.}~\bibnamefont
  {Giacovazzo}},\ }\href {https://books.google.com/books?id=6oAfAgAAQBAJ}
  {\emph {\bibinfo {title} {Phasing in Crystallography: A Modern
  Perspective}}},\ International Union of Crystallography Texts on
  Crystallography\ (\bibinfo  {publisher} {OUP Oxford},\ \bibinfo {year}
  {2014})\BibitemShut {NoStop}%
\bibitem [{\citenamefont {Snieder}\ and\ \citenamefont
  {Larose}(2013)}]{Snieder+2013}%
  \BibitemOpen
  \bibfield  {author} {\bibinfo {author} {\bibfnamefont {R.}~\bibnamefont
  {Snieder}}\ and\ \bibinfo {author} {\bibfnamefont {E.}~\bibnamefont
  {Larose}},\ }\bibfield  {title} {\bibinfo {title} {Extracting earth's elastic
  wave response from noise measurements},\ }\href
  {https://doi.org/10.1146/annurev-earth-050212-123936} {\bibfield  {journal}
  {\bibinfo  {journal} {Annual Review of Earth and Planetary Sciences}\
  }\textbf {\bibinfo {volume} {41}},\ \bibinfo {pages} {183} (\bibinfo {year}
  {2013})},\ \Eprint
  {https://arxiv.org/abs/https://doi.org/10.1146/annurev-earth-050212-123936}
  {https://doi.org/10.1146/annurev-earth-050212-123936} \BibitemShut {NoStop}%
\bibitem [{\citenamefont {Callow}(2003)}]{Callow2003}%
  \BibitemOpen
  \bibfield  {author} {\bibinfo {author} {\bibfnamefont {H.}~\bibnamefont
  {Callow}},\ }\href {https://books.google.com/books?id=j0K0MwAACAAJ} {\emph
  {\bibinfo {title} {Signal Processing for Synthetic Aperture Sonar Image
  Enhancement: A Thesis Presented for the Degree of Doctor of Philosophy in
  Electrical and Electronic Engineering at the University of Canterbury,
  Christchurch, New Zealand}}}\ (\bibinfo  {publisher} {University of
  Canterbury},\ \bibinfo {year} {2003})\BibitemShut {NoStop}%
\bibitem [{\citenamefont {Rosen}\ \emph {et~al.}(2000)\citenamefont {Rosen},
  \citenamefont {Hensley}, \citenamefont {Joughin}, \citenamefont {Li},
  \citenamefont {Madsen}, \citenamefont {Rodriguez},\ and\ \citenamefont
  {Goldstein}}]{Rosen+2000}%
  \BibitemOpen
  \bibfield  {author} {\bibinfo {author} {\bibfnamefont {P.}~\bibnamefont
  {Rosen}}, \bibinfo {author} {\bibfnamefont {S.}~\bibnamefont {Hensley}},
  \bibinfo {author} {\bibfnamefont {I.}~\bibnamefont {Joughin}}, \bibinfo
  {author} {\bibfnamefont {F.}~\bibnamefont {Li}}, \bibinfo {author}
  {\bibfnamefont {S.}~\bibnamefont {Madsen}}, \bibinfo {author} {\bibfnamefont
  {E.}~\bibnamefont {Rodriguez}},\ and\ \bibinfo {author} {\bibfnamefont
  {R.}~\bibnamefont {Goldstein}},\ }\bibfield  {title} {\bibinfo {title}
  {Synthetic aperture radar interferometry},\ }\href
  {https://doi.org/10.1109/5.838084} {\bibfield  {journal} {\bibinfo  {journal}
  {Proceedings of the IEEE}\ }\textbf {\bibinfo {volume} {88}},\ \bibinfo
  {pages} {333 } (\bibinfo {year} {2000})}\BibitemShut {NoStop}%
\bibitem [{\citenamefont {{Thompson}}\ \emph {et~al.}(2017)\citenamefont
  {{Thompson}}, \citenamefont {{Moran}},\ and\ \citenamefont
  {{Swenson}}}]{TMS2017}%
  \BibitemOpen
  \bibfield  {author} {\bibinfo {author} {\bibfnamefont {A.~R.}\ \bibnamefont
  {{Thompson}}}, \bibinfo {author} {\bibfnamefont {J.~M.}\ \bibnamefont
  {{Moran}}},\ and\ \bibinfo {author} {\bibfnamefont {J.}~\bibnamefont
  {{Swenson}}, \bibfnamefont {George~W.}},\ }\href
  {https://doi.org/10.1007/978-3-319-44431-4} {\emph {\bibinfo {title}
  {{Interferometry and Synthesis in Radio Astronomy, 3rd Edition}}}}\ (\bibinfo
   {publisher} {Springer, Cham},\ \bibinfo {year} {2017})\BibitemShut {NoStop}%
\bibitem [{\citenamefont {{Taylor}}\ \emph {et~al.}(1999)\citenamefont
  {{Taylor}}, \citenamefont {{Carilli}},\ and\ \citenamefont
  {{Perley}}}]{SIRA-II}%
  \BibitemOpen
  \bibinfo {editor} {\bibfnamefont {G.~B.}\ \bibnamefont {{Taylor}}}, \bibinfo
  {editor} {\bibfnamefont {C.~L.}\ \bibnamefont {{Carilli}}},\ and\ \bibinfo
  {editor} {\bibfnamefont {R.~A.}\ \bibnamefont {{Perley}}},\ eds.,\ \href@noop
  {} {\emph {\bibinfo {title} {Synthesis Imaging in Radio Astronomy II}}},\
  \bibinfo {series} {Astronomical Society of the Pacific Conference Series},
  Vol.\ \bibinfo {volume} {180}\ (\bibinfo  {publisher} {San Francisco, Calif.
  : Astronomical Society of the Pacific},\ \bibinfo {year} {1999})\BibitemShut
  {NoStop}%
\bibitem [{\citenamefont {{Cornwell}}\ and\ \citenamefont
  {{Wilkinson}}(1981)}]{Cornwell+1981}%
  \BibitemOpen
  \bibfield  {author} {\bibinfo {author} {\bibfnamefont {T.~J.}\ \bibnamefont
  {{Cornwell}}}\ and\ \bibinfo {author} {\bibfnamefont {P.~N.}\ \bibnamefont
  {{Wilkinson}}},\ }\bibfield  {title} {\bibinfo {title} {{A new method for
  making maps with unstable radio interferometers}},\ }\href
  {https://doi.org/10.1093/mnras/196.4.1067} {\bibfield  {journal} {\bibinfo
  {journal} {\mnras}\ }\textbf {\bibinfo {volume} {196}},\ \bibinfo {pages}
  {1067} (\bibinfo {year} {1981})}\BibitemShut {NoStop}%
\bibitem [{\citenamefont {{Ekers}}(1984)}]{Ekers1984}%
  \BibitemOpen
  \bibfield  {author} {\bibinfo {author} {\bibfnamefont {R.~D.}\ \bibnamefont
  {{Ekers}}},\ }\bibfield  {title} {\bibinfo {title} {{The almost Serendipitous
  Discovery of Self-Calibration}},\ }in\ \href@noop {} {\emph {\bibinfo
  {booktitle} {Serendipitous Discoveries in Radio Astronomy}}}\ (\bibinfo
  {year} {1984})\ p.\ \bibinfo {pages} {154}\BibitemShut {NoStop}%
\bibitem [{\citenamefont {{Monnier}}(2007)}]{mon07b}%
  \BibitemOpen
  \bibfield  {author} {\bibinfo {author} {\bibfnamefont {J.~D.}\ \bibnamefont
  {{Monnier}}},\ }\bibfield  {title} {\bibinfo {title} {{Phases in
  interferometry}},\ }\href {https://doi.org/10.1016/j.newar.2007.06.006}
  {\bibfield  {journal} {\bibinfo  {journal} {\nar}\ }\textbf {\bibinfo
  {volume} {51}},\ \bibinfo {pages} {604} (\bibinfo {year} {2007})}\BibitemShut
  {NoStop}%
\bibitem [{\citenamefont {{Bargmann}}(1964)}]{Bargmann1964}%
  \BibitemOpen
  \bibfield  {author} {\bibinfo {author} {\bibfnamefont {V.}~\bibnamefont
  {{Bargmann}}},\ }\bibfield  {title} {\bibinfo {title} {{Note on Wigner's
  Theorem on Symmetry Operations}},\ }\href {https://doi.org/10.1063/1.1704188}
  {\bibfield  {journal} {\bibinfo  {journal} {Journal of Mathematical Physics}\
  }\textbf {\bibinfo {volume} {5}},\ \bibinfo {pages} {862} (\bibinfo {year}
  {1964})}\BibitemShut {NoStop}%
\bibitem [{\citenamefont {{Thyagarajan}}\ and\ \citenamefont
  {{Carilli}}(2020)}]{Thyagarajan+2020c}%
  \BibitemOpen
  \bibfield  {author} {\bibinfo {author} {\bibfnamefont {N.}~\bibnamefont
  {{Thyagarajan}}}\ and\ \bibinfo {author} {\bibfnamefont {C.~L.}\ \bibnamefont
  {{Carilli}}},\ }\bibfield  {title} {\bibinfo {title} {{Invariants in
  Interferometry: Geometric Insight into Closure Phases}},\ }\href@noop {}
  {\bibfield  {journal} {\bibinfo  {journal} {arXiv e-prints}\ ,\ \bibinfo
  {eid} {arXiv:2012.05254}} (\bibinfo {year} {2020})},\ \Eprint
  {https://arxiv.org/abs/2012.05254} {arXiv:2012.05254 [astro-ph.IM]}
  \BibitemShut {NoStop}%
\bibitem [{\citenamefont {{Jennison}}(1958)}]{jen58}%
  \BibitemOpen
  \bibfield  {author} {\bibinfo {author} {\bibfnamefont {R.~C.}\ \bibnamefont
  {{Jennison}}},\ }\bibfield  {title} {\bibinfo {title} {{A phase sensitive
  interferometer technique for the measurement of the Fourier transforms of
  spatial brightness distributions of small angular extent}},\ }\href
  {https://doi.org/10.1093/mnras/118.3.276} {\bibfield  {journal} {\bibinfo
  {journal} {\mnras}\ }\textbf {\bibinfo {volume} {118}},\ \bibinfo {pages}
  {276} (\bibinfo {year} {1958})}\BibitemShut {NoStop}%
\bibitem [{\citenamefont {{Twiss}}\ \emph {et~al.}(1960)\citenamefont
  {{Twiss}}, \citenamefont {{Carter}},\ and\ \citenamefont
  {{Little}}}]{Twiss+1960}%
  \BibitemOpen
  \bibfield  {author} {\bibinfo {author} {\bibfnamefont {R.~Q.}\ \bibnamefont
  {{Twiss}}}, \bibinfo {author} {\bibfnamefont {A.~W.~L.}\ \bibnamefont
  {{Carter}}},\ and\ \bibinfo {author} {\bibfnamefont {A.~G.}\ \bibnamefont
  {{Little}}},\ }\bibfield  {title} {\bibinfo {title} {{Brightness distribution
  over some strong radio sources at 1427 Mc/s}},\ }\href@noop {} {\bibfield
  {journal} {\bibinfo  {journal} {The Observatory}\ }\textbf {\bibinfo {volume}
  {80}},\ \bibinfo {pages} {153} (\bibinfo {year} {1960})}\BibitemShut
  {NoStop}%
\bibitem [{\citenamefont {{Monnier}}(2003{\natexlab{a}})}]{mon03a}%
  \BibitemOpen
  \bibfield  {author} {\bibinfo {author} {\bibfnamefont {J.~D.}\ \bibnamefont
  {{Monnier}}},\ }\bibfield  {title} {\bibinfo {title} {{Astrophysics with
  Closure Phases}},\ }in\ \href {https://doi.org/10.1051/eas:2003019} {\emph
  {\bibinfo {booktitle} {EAS Publications Series}}},\ \bibinfo {series} {EAS
  Publications Series}, Vol.~\bibinfo {volume} {6},\ \bibinfo {editor} {edited
  by\ \bibinfo {editor} {\bibfnamefont {G.}~\bibnamefont {{Perrin}}}\ and\
  \bibinfo {editor} {\bibfnamefont {F.}~\bibnamefont {{Malbet}}}}\ (\bibinfo
  {year} {2003})\ p.\ \bibinfo {pages} {213}\BibitemShut {NoStop}%
\bibitem [{\citenamefont {{Monnier}}(2003{\natexlab{b}})}]{mon03b}%
  \BibitemOpen
  \bibfield  {author} {\bibinfo {author} {\bibfnamefont {J.~D.}\ \bibnamefont
  {{Monnier}}},\ }\bibfield  {title} {\bibinfo {title} {{Optical interferometry
  in astronomy}},\ }\href {https://doi.org/10.1088/0034-4885/66/5/203}
  {\bibfield  {journal} {\bibinfo  {journal} {Reports on Progress in Physics}\
  }\textbf {\bibinfo {volume} {66}},\ \bibinfo {pages} {789} (\bibinfo {year}
  {2003}{\natexlab{b}})},\ \Eprint {https://arxiv.org/abs/astro-ph/0307036}
  {arXiv:astro-ph/0307036 [astro-ph]} \BibitemShut {NoStop}%
\bibitem [{\citenamefont {{Monnier}}\ \emph {et~al.}(2006)\citenamefont
  {{Monnier}}, \citenamefont {{Berger}}, \citenamefont {{Millan-Gabet}},
  \citenamefont {{Traub}}, \citenamefont {{Schloerb}}, \citenamefont
  {{Pedretti}}, \citenamefont {{Benisty}}, \citenamefont {{Carleton}},
  \citenamefont {{Haguenauer}}, \citenamefont {{Kern}}, \citenamefont
  {{Labeye}}, \citenamefont {{Lacasse}}, \citenamefont {{Malbet}},
  \citenamefont {{Perraut}}, \citenamefont {{Pearlman}},\ and\ \citenamefont
  {{Zhao}}}]{mon06}%
  \BibitemOpen
  \bibfield  {author} {\bibinfo {author} {\bibfnamefont {J.~D.}\ \bibnamefont
  {{Monnier}}}, \bibinfo {author} {\bibfnamefont {J.-P.}\ \bibnamefont
  {{Berger}}}, \bibinfo {author} {\bibfnamefont {R.}~\bibnamefont
  {{Millan-Gabet}}}, \bibinfo {author} {\bibfnamefont {W.~A.}\ \bibnamefont
  {{Traub}}}, \bibinfo {author} {\bibfnamefont {F.~P.}\ \bibnamefont
  {{Schloerb}}}, \bibinfo {author} {\bibfnamefont {E.}~\bibnamefont
  {{Pedretti}}}, \bibinfo {author} {\bibfnamefont {M.}~\bibnamefont
  {{Benisty}}}, \bibinfo {author} {\bibfnamefont {N.~P.}\ \bibnamefont
  {{Carleton}}}, \bibinfo {author} {\bibfnamefont {P.}~\bibnamefont
  {{Haguenauer}}}, \bibinfo {author} {\bibfnamefont {P.}~\bibnamefont
  {{Kern}}}, \bibinfo {author} {\bibfnamefont {P.}~\bibnamefont {{Labeye}}},
  \bibinfo {author} {\bibfnamefont {M.~G.}\ \bibnamefont {{Lacasse}}}, \bibinfo
  {author} {\bibfnamefont {F.}~\bibnamefont {{Malbet}}}, \bibinfo {author}
  {\bibfnamefont {K.}~\bibnamefont {{Perraut}}}, \bibinfo {author}
  {\bibfnamefont {M.}~\bibnamefont {{Pearlman}}},\ and\ \bibinfo {author}
  {\bibfnamefont {M.}~\bibnamefont {{Zhao}}},\ }\bibfield  {title} {\bibinfo
  {title} {{Few Skewed Disks Found in First Closure-Phase Survey of Herbig
  Ae/Be Stars}},\ }\href {https://doi.org/10.1086/505340} {\bibfield  {journal}
  {\bibinfo  {journal} {\apj}\ }\textbf {\bibinfo {volume} {647}},\ \bibinfo
  {pages} {444} (\bibinfo {year} {2006})},\ \Eprint
  {https://arxiv.org/abs/astro-ph/0606052} {astro-ph/0606052} \BibitemShut
  {NoStop}%
\bibitem [{\citenamefont {{Monnier}}\ \emph {et~al.}(2007)\citenamefont
  {{Monnier}}, \citenamefont {{Zhao}}, \citenamefont {{Pedretti}},
  \citenamefont {{Thureau}}, \citenamefont {{Ireland}}, \citenamefont
  {{Muirhead}}, \citenamefont {{Berger}}, \citenamefont {{Millan-Gabet}},
  \citenamefont {{Van Belle}}, \citenamefont {{ten Brummelaar}}, \citenamefont
  {{McAlister}}, \citenamefont {{Ridgway}}, \citenamefont {{Turner}},
  \citenamefont {{Sturmann}}, \citenamefont {{Sturmann}},\ and\ \citenamefont
  {{Berger}}}]{mon07a}%
  \BibitemOpen
  \bibfield  {author} {\bibinfo {author} {\bibfnamefont {J.~D.}\ \bibnamefont
  {{Monnier}}}, \bibinfo {author} {\bibfnamefont {M.}~\bibnamefont {{Zhao}}},
  \bibinfo {author} {\bibfnamefont {E.}~\bibnamefont {{Pedretti}}}, \bibinfo
  {author} {\bibfnamefont {N.}~\bibnamefont {{Thureau}}}, \bibinfo {author}
  {\bibfnamefont {M.}~\bibnamefont {{Ireland}}}, \bibinfo {author}
  {\bibfnamefont {P.}~\bibnamefont {{Muirhead}}}, \bibinfo {author}
  {\bibfnamefont {J.~P.}\ \bibnamefont {{Berger}}}, \bibinfo {author}
  {\bibfnamefont {R.}~\bibnamefont {{Millan-Gabet}}}, \bibinfo {author}
  {\bibfnamefont {G.}~\bibnamefont {{Van Belle}}}, \bibinfo {author}
  {\bibfnamefont {T.}~\bibnamefont {{ten Brummelaar}}}, \bibinfo {author}
  {\bibfnamefont {H.}~\bibnamefont {{McAlister}}}, \bibinfo {author}
  {\bibfnamefont {S.}~\bibnamefont {{Ridgway}}}, \bibinfo {author}
  {\bibfnamefont {N.}~\bibnamefont {{Turner}}}, \bibinfo {author}
  {\bibfnamefont {L.}~\bibnamefont {{Sturmann}}}, \bibinfo {author}
  {\bibfnamefont {J.}~\bibnamefont {{Sturmann}}},\ and\ \bibinfo {author}
  {\bibfnamefont {D.}~\bibnamefont {{Berger}}},\ }\bibfield  {title} {\bibinfo
  {title} {{Imaging the Surface of Altair}},\ }\href
  {https://doi.org/10.1126/science.1143205} {\bibfield  {journal} {\bibinfo
  {journal} {Science}\ }\textbf {\bibinfo {volume} {317}},\ \bibinfo {pages}
  {342} (\bibinfo {year} {2007})},\ \Eprint {https://arxiv.org/abs/0706.0867}
  {arXiv:0706.0867 [astro-ph]} \BibitemShut {NoStop}%
\bibitem [{\citenamefont {Schwab}(1980)}]{Schwab1980}%
  \BibitemOpen
  \bibfield  {author} {\bibinfo {author} {\bibfnamefont {F.~R.}\ \bibnamefont
  {Schwab}},\ }\bibfield  {title} {\bibinfo {title} {{Adaptive Calibration Of
  Radio Interferometer Data}},\ }in\ \href {https://doi.org/10.1117/12.958828}
  {\emph {\bibinfo {booktitle} {1980 Intl Optical Computing Conf I}}},\ Vol.\
  \bibinfo {volume} {0231},\ \bibinfo {editor} {edited by\ \bibinfo {editor}
  {\bibfnamefont {W.~T.}\ \bibnamefont {Rhodes}}},\ \bibinfo {organization}
  {International Society for Optics and Photonics}\ (\bibinfo  {publisher}
  {SPIE},\ \bibinfo {year} {1980})\ pp.\ \bibinfo {pages} {18 --
  25}\BibitemShut {NoStop}%
\bibitem [{\citenamefont {{Pearson}}\ and\ \citenamefont
  {{Readhead}}(1984)}]{Pearson+1984}%
  \BibitemOpen
  \bibfield  {author} {\bibinfo {author} {\bibfnamefont {T.~J.}\ \bibnamefont
  {{Pearson}}}\ and\ \bibinfo {author} {\bibfnamefont {A.~C.~S.}\ \bibnamefont
  {{Readhead}}},\ }\bibfield  {title} {\bibinfo {title} {{Image Formation by
  Self-Calibration in Radio Astronomy}},\ }\href
  {https://doi.org/10.1146/annurev.aa.22.090184.000525} {\bibfield  {journal}
  {\bibinfo  {journal} {\araa}\ }\textbf {\bibinfo {volume} {22}},\ \bibinfo
  {pages} {97} (\bibinfo {year} {1984})}\BibitemShut {NoStop}%
\bibitem [{\citenamefont {{Bhatnagar}}\ and\ \citenamefont
  {{Nityananda}}(2001)}]{bhatnagar+2001}%
  \BibitemOpen
  \bibfield  {author} {\bibinfo {author} {\bibfnamefont {S.}~\bibnamefont
  {{Bhatnagar}}}\ and\ \bibinfo {author} {\bibfnamefont {R.}~\bibnamefont
  {{Nityananda}}},\ }\bibfield  {title} {\bibinfo {title} {{Solving for closure
  errors due to polarization leakage in radio interferometry of unpolarized
  sources}},\ }\href {https://doi.org/10.1051/0004-6361:20010799} {\bibfield
  {journal} {\bibinfo  {journal} {\aap}\ }\textbf {\bibinfo {volume} {375}},\
  \bibinfo {pages} {344} (\bibinfo {year} {2001})},\ \Eprint
  {https://arxiv.org/abs/astro-ph/0106348} {arXiv:astro-ph/0106348 [astro-ph]}
  \BibitemShut {NoStop}%
\bibitem [{\citenamefont {{Thyagarajan}}\ \emph {et~al.}(2018)\citenamefont
  {{Thyagarajan}}, \citenamefont {{Carilli}},\ and\ \citenamefont
  {{Nikolic}}}]{thy18}%
  \BibitemOpen
  \bibfield  {author} {\bibinfo {author} {\bibfnamefont {N.}~\bibnamefont
  {{Thyagarajan}}}, \bibinfo {author} {\bibfnamefont {C.~L.}\ \bibnamefont
  {{Carilli}}},\ and\ \bibinfo {author} {\bibfnamefont {B.}~\bibnamefont
  {{Nikolic}}},\ }\bibfield  {title} {\bibinfo {title} {{Detecting Cosmic
  Reionization Using the Bispectrum Phase}},\ }\href
  {https://doi.org/10.1103/PhysRevLett.120.251301} {\bibfield  {journal}
  {\bibinfo  {journal} {Physical Review Letters}\ }\textbf {\bibinfo {volume}
  {120}},\ \bibinfo {eid} {251301} (\bibinfo {year} {2018})},\ \Eprint
  {https://arxiv.org/abs/1805.00954} {arXiv:1805.00954} \BibitemShut {NoStop}%
\bibitem [{\citenamefont {Thyagarajan}\ and\ \citenamefont
  {Carilli}(2020)}]{thy20a}%
  \BibitemOpen
  \bibfield  {author} {\bibinfo {author} {\bibfnamefont {N.}~\bibnamefont
  {Thyagarajan}}\ and\ \bibinfo {author} {\bibfnamefont {C.~L.}\ \bibnamefont
  {Carilli}},\ }\bibfield  {title} {\bibinfo {title} {Detection of cosmic
  structures using the bispectrum phase. i. mathematical foundations},\ }\href
  {https://doi.org/10.1103/PhysRevD.102.022001} {\bibfield  {journal} {\bibinfo
   {journal} {Phys. Rev. D}\ }\textbf {\bibinfo {volume} {102}},\ \bibinfo
  {pages} {022001} (\bibinfo {year} {2020})},\ \Eprint
  {https://arxiv.org/abs/2005.10274} {arXiv:2005.10274} \BibitemShut {NoStop}%
\bibitem [{\citenamefont {Thyagarajan}\ \emph {et~al.}(2020)\citenamefont
  {Thyagarajan}, \citenamefont {Carilli}, \citenamefont {Nikolic},
  \citenamefont {Kent}, \citenamefont {Mesinger}, \citenamefont {Kern},
  \citenamefont {Bernardi}, \citenamefont {Matika}, \citenamefont
  {Abdurashidova}, \citenamefont {Aguirre}, \citenamefont {Alexander},
  \citenamefont {Ali}, \citenamefont {Balfour}, \citenamefont {Beardsley},
  \citenamefont {Billings}, \citenamefont {Bowman}, \citenamefont {Bradley},
  \citenamefont {Burba}, \citenamefont {Carey}, \citenamefont {Cheng},
  \citenamefont {DeBoer}, \citenamefont {Dexter}, \citenamefont {Acedo},
  \citenamefont {Dillon}, \citenamefont {Ely}, \citenamefont {Ewall-Wice},
  \citenamefont {Fagnoni}, \citenamefont {Fritz}, \citenamefont {Furlanetto},
  \citenamefont {Gale-Sides}, \citenamefont {Glendenning}, \citenamefont
  {Gorthi}, \citenamefont {Greig}, \citenamefont {Grobbelaar}, \citenamefont
  {Halday}, \citenamefont {Hazelton}, \citenamefont {Hewitt}, \citenamefont
  {Hickish}, \citenamefont {Jacobs}, \citenamefont {Julius}, \citenamefont
  {Kerrigan}, \citenamefont {Kittiwisit}, \citenamefont {Kohn}, \citenamefont
  {Kolopanis}, \citenamefont {Lanman}, \citenamefont {La~Plante}, \citenamefont
  {Lekalake}, \citenamefont {Lewis}, \citenamefont {Liu}, \citenamefont
  {MacMahon}, \citenamefont {Malan}, \citenamefont {Malgas}, \citenamefont
  {Maree}, \citenamefont {Martinot}, \citenamefont {Matsetela}, \citenamefont
  {Molewa}, \citenamefont {Morales}, \citenamefont {Mosiane}, \citenamefont
  {Neben}, \citenamefont {Parsons}, \citenamefont {Patra}, \citenamefont
  {Pieterse}, \citenamefont {Pober}, \citenamefont {Razavi-Ghods},
  \citenamefont {Ringuette}, \citenamefont {Robnett}, \citenamefont {Rosie},
  \citenamefont {Sims}, \citenamefont {Smith}, \citenamefont {Syce},
  \citenamefont {Williams},\ and\ \citenamefont {Zheng}}]{thy20b}%
  \BibitemOpen
  \bibfield  {author} {\bibinfo {author} {\bibfnamefont {N.}~\bibnamefont
  {Thyagarajan}}, \bibinfo {author} {\bibfnamefont {C.~L.}\ \bibnamefont
  {Carilli}}, \bibinfo {author} {\bibfnamefont {B.}~\bibnamefont {Nikolic}},
  \bibinfo {author} {\bibfnamefont {J.}~\bibnamefont {Kent}}, \bibinfo {author}
  {\bibfnamefont {A.}~\bibnamefont {Mesinger}}, \bibinfo {author}
  {\bibfnamefont {N.~S.}\ \bibnamefont {Kern}}, \bibinfo {author}
  {\bibfnamefont {G.}~\bibnamefont {Bernardi}}, \bibinfo {author}
  {\bibfnamefont {S.}~\bibnamefont {Matika}}, \bibinfo {author} {\bibfnamefont
  {Z.}~\bibnamefont {Abdurashidova}}, \bibinfo {author} {\bibfnamefont {J.~E.}\
  \bibnamefont {Aguirre}}, \bibinfo {author} {\bibfnamefont {P.}~\bibnamefont
  {Alexander}}, \bibinfo {author} {\bibfnamefont {Z.~S.}\ \bibnamefont {Ali}},
  \bibinfo {author} {\bibfnamefont {Y.}~\bibnamefont {Balfour}}, \bibinfo
  {author} {\bibfnamefont {A.~P.}\ \bibnamefont {Beardsley}}, \bibinfo {author}
  {\bibfnamefont {T.~S.}\ \bibnamefont {Billings}}, \bibinfo {author}
  {\bibfnamefont {J.~D.}\ \bibnamefont {Bowman}}, \bibinfo {author}
  {\bibfnamefont {R.~F.}\ \bibnamefont {Bradley}}, \bibinfo {author}
  {\bibfnamefont {J.}~\bibnamefont {Burba}}, \bibinfo {author} {\bibfnamefont
  {S.}~\bibnamefont {Carey}}, \bibinfo {author} {\bibfnamefont
  {C.}~\bibnamefont {Cheng}}, \bibinfo {author} {\bibfnamefont {D.~R.}\
  \bibnamefont {DeBoer}}, \bibinfo {author} {\bibfnamefont {M.}~\bibnamefont
  {Dexter}}, \bibinfo {author} {\bibfnamefont {E.~d.~L.}\ \bibnamefont
  {Acedo}}, \bibinfo {author} {\bibfnamefont {J.~S.}\ \bibnamefont {Dillon}},
  \bibinfo {author} {\bibfnamefont {J.}~\bibnamefont {Ely}}, \bibinfo {author}
  {\bibfnamefont {A.}~\bibnamefont {Ewall-Wice}}, \bibinfo {author}
  {\bibfnamefont {N.}~\bibnamefont {Fagnoni}}, \bibinfo {author} {\bibfnamefont
  {R.}~\bibnamefont {Fritz}}, \bibinfo {author} {\bibfnamefont {S.~R.}\
  \bibnamefont {Furlanetto}}, \bibinfo {author} {\bibfnamefont
  {K.}~\bibnamefont {Gale-Sides}}, \bibinfo {author} {\bibfnamefont
  {B.}~\bibnamefont {Glendenning}}, \bibinfo {author} {\bibfnamefont
  {D.}~\bibnamefont {Gorthi}}, \bibinfo {author} {\bibfnamefont
  {B.}~\bibnamefont {Greig}}, \bibinfo {author} {\bibfnamefont
  {J.}~\bibnamefont {Grobbelaar}}, \bibinfo {author} {\bibfnamefont
  {Z.}~\bibnamefont {Halday}}, \bibinfo {author} {\bibfnamefont {B.~J.}\
  \bibnamefont {Hazelton}}, \bibinfo {author} {\bibfnamefont {J.~N.}\
  \bibnamefont {Hewitt}}, \bibinfo {author} {\bibfnamefont {J.}~\bibnamefont
  {Hickish}}, \bibinfo {author} {\bibfnamefont {D.~C.}\ \bibnamefont {Jacobs}},
  \bibinfo {author} {\bibfnamefont {A.}~\bibnamefont {Julius}}, \bibinfo
  {author} {\bibfnamefont {J.}~\bibnamefont {Kerrigan}}, \bibinfo {author}
  {\bibfnamefont {P.}~\bibnamefont {Kittiwisit}}, \bibinfo {author}
  {\bibfnamefont {S.~A.}\ \bibnamefont {Kohn}}, \bibinfo {author}
  {\bibfnamefont {M.}~\bibnamefont {Kolopanis}}, \bibinfo {author}
  {\bibfnamefont {A.}~\bibnamefont {Lanman}}, \bibinfo {author} {\bibfnamefont
  {P.}~\bibnamefont {La~Plante}}, \bibinfo {author} {\bibfnamefont
  {T.}~\bibnamefont {Lekalake}}, \bibinfo {author} {\bibfnamefont
  {D.}~\bibnamefont {Lewis}}, \bibinfo {author} {\bibfnamefont
  {A.}~\bibnamefont {Liu}}, \bibinfo {author} {\bibfnamefont {D.}~\bibnamefont
  {MacMahon}}, \bibinfo {author} {\bibfnamefont {L.}~\bibnamefont {Malan}},
  \bibinfo {author} {\bibfnamefont {C.}~\bibnamefont {Malgas}}, \bibinfo
  {author} {\bibfnamefont {M.}~\bibnamefont {Maree}}, \bibinfo {author}
  {\bibfnamefont {Z.~E.}\ \bibnamefont {Martinot}}, \bibinfo {author}
  {\bibfnamefont {E.}~\bibnamefont {Matsetela}}, \bibinfo {author}
  {\bibfnamefont {M.}~\bibnamefont {Molewa}}, \bibinfo {author} {\bibfnamefont
  {M.~F.}\ \bibnamefont {Morales}}, \bibinfo {author} {\bibfnamefont
  {T.}~\bibnamefont {Mosiane}}, \bibinfo {author} {\bibfnamefont {A.~R.}\
  \bibnamefont {Neben}}, \bibinfo {author} {\bibfnamefont {A.~R.}\ \bibnamefont
  {Parsons}}, \bibinfo {author} {\bibfnamefont {N.}~\bibnamefont {Patra}},
  \bibinfo {author} {\bibfnamefont {S.}~\bibnamefont {Pieterse}}, \bibinfo
  {author} {\bibfnamefont {J.~C.}\ \bibnamefont {Pober}}, \bibinfo {author}
  {\bibfnamefont {N.}~\bibnamefont {Razavi-Ghods}}, \bibinfo {author}
  {\bibfnamefont {J.}~\bibnamefont {Ringuette}}, \bibinfo {author}
  {\bibfnamefont {J.}~\bibnamefont {Robnett}}, \bibinfo {author} {\bibfnamefont
  {K.}~\bibnamefont {Rosie}}, \bibinfo {author} {\bibfnamefont
  {P.}~\bibnamefont {Sims}}, \bibinfo {author} {\bibfnamefont {C.}~\bibnamefont
  {Smith}}, \bibinfo {author} {\bibfnamefont {A.}~\bibnamefont {Syce}},
  \bibinfo {author} {\bibfnamefont {P.~K.~G.}\ \bibnamefont {Williams}},\ and\
  \bibinfo {author} {\bibfnamefont {H.}~\bibnamefont {Zheng}},\ }\bibfield
  {title} {\bibinfo {title} {Detection of cosmic structures using the
  bispectrum phase. ii. first results from application to cosmic reionization
  using the hydrogen epoch of reionization array},\ }\href
  {https://doi.org/10.1103/PhysRevD.102.022002} {\bibfield  {journal} {\bibinfo
   {journal} {Phys. Rev. D}\ }\textbf {\bibinfo {volume} {102}},\ \bibinfo
  {pages} {022002} (\bibinfo {year} {2020})},\ \Eprint
  {https://arxiv.org/abs/2005.10275} {arXiv:2005.10275 [astro-ph.CO]}
  \BibitemShut {NoStop}%
\bibitem [{\citenamefont {{Lannes}}(1991)}]{Lannes1991}%
  \BibitemOpen
  \bibfield  {author} {\bibinfo {author} {\bibfnamefont {A.}~\bibnamefont
  {{Lannes}}},\ }\bibfield  {title} {\bibinfo {title} {{Phase and amplitude
  calibration in aperture synthesis. Algebraic structures}},\ }\href
  {https://doi.org/10.1088/0266-5611/7/2/009} {\bibfield  {journal} {\bibinfo
  {journal} {Inverse Problems}\ }\textbf {\bibinfo {volume} {7}},\ \bibinfo
  {pages} {261} (\bibinfo {year} {1991})}\BibitemShut {NoStop}%
\bibitem [{\citenamefont {{Broderick}}\ and\ \citenamefont
  {{Pesce}}(2020)}]{Broderick+2020}%
  \BibitemOpen
  \bibfield  {author} {\bibinfo {author} {\bibfnamefont {A.~E.}\ \bibnamefont
  {{Broderick}}}\ and\ \bibinfo {author} {\bibfnamefont {D.~W.}\ \bibnamefont
  {{Pesce}}},\ }\bibfield  {title} {\bibinfo {title} {{Closure Traces: Novel
  Calibration-insensitive Quantities for Radio Astronomy}},\ }\href
  {https://doi.org/10.3847/1538-4357/abbd9d} {\bibfield  {journal} {\bibinfo
  {journal} {\apj}\ }\textbf {\bibinfo {volume} {904}},\ \bibinfo {eid} {126}
  (\bibinfo {year} {2020})},\ \Eprint {https://arxiv.org/abs/2010.00612}
  {arXiv:2010.00612 [astro-ph.IM]} \BibitemShut {NoStop}%
\bibitem [{\citenamefont {Samuel}\ \emph {et~al.}(2021)\citenamefont {Samuel},
  \citenamefont {Nityananda},\ and\ \citenamefont
  {Thyagarajan}}]{polarimetric-invariants}%
  \BibitemOpen
  \bibfield  {author} {\bibinfo {author} {\bibfnamefont {J.}~\bibnamefont
  {Samuel}}, \bibinfo {author} {\bibfnamefont {R.}~\bibnamefont {Nityananda}},\
  and\ \bibinfo {author} {\bibfnamefont {N.}~\bibnamefont {Thyagarajan}},\
  }\href@noop {} {\bibinfo {title} {Invariants in polarimetric interferometry:
  a non-abelian gauge theory}} (\bibinfo {year} {2021}),\ \Eprint
  {https://arxiv.org/abs/2108.11400} {arXiv:2108.11400 [gr-qc]} \BibitemShut
  {NoStop}%
\bibitem [{\citenamefont {{Pancharatnam}}(1956)}]{Pancharatnam1956a}%
  \BibitemOpen
  \bibfield  {author} {\bibinfo {author} {\bibfnamefont {S.}~\bibnamefont
  {{Pancharatnam}}},\ }\bibfield  {title} {\bibinfo {title} {{Generalized
  theory of interference, and its applications. Part I. Coherent pencils}},\
  }\href {https://doi.org/10.1007/BF03046050} {\bibfield  {journal} {\bibinfo
  {journal} {Proceedings of the Indian Academy of Sciences}\ }\textbf {\bibinfo
  {volume} {A44}},\ \bibinfo {pages} {247} (\bibinfo {year}
  {1956})}\BibitemShut {NoStop}%
\bibitem [{\citenamefont {{Pancharatnam}}(1976)}]{Pancharatnam1975}%
  \BibitemOpen
  \bibfield  {author} {\bibinfo {author} {\bibfnamefont {S.}~\bibnamefont
  {{Pancharatnam}}},\ }\bibfield  {title} {\bibinfo {title} {Collected works of
  {S}. {Pancharatnam}},\ }\href {https://doi.org/10.1088/0031-9112/27/6/028}
  {\bibfield  {journal} {\bibinfo  {journal} {Physics Bulletin}\ }\textbf
  {\bibinfo {volume} {27}},\ \bibinfo {pages} {265} (\bibinfo {year}
  {1976})}\BibitemShut {NoStop}%
\bibitem [{\citenamefont {Klein}(1872)}]{Klein1872}%
  \BibitemOpen
  \bibfield  {author} {\bibinfo {author} {\bibfnamefont {F.}~\bibnamefont
  {Klein}},\ }\href@noop {} {\emph {\bibinfo {title} {Vergleichende
  Betrachtungen \"uber neuere geometrische Forschungen}}}\ (\bibinfo
  {publisher} {Verlag von Andreas Deichert},\ \bibinfo {address} {Erlangen},\
  \bibinfo {year} {1872})\ \bibinfo {note} {published later in
  \emph{Mathematische Annalen} 43 (1893), 63--100. English translation by M. W.
  Haskell, A comparative review of recent researches in geometry,
  \emph{Bulletin of the New York Mathematical Society} 2 (1892--1893),
  215--249.}\BibitemShut {Stop}%
\bibitem [{\citenamefont {{Penrose}}\ and\ \citenamefont
  {{Rindler}}(1984)}]{Penrose+1984}%
  \BibitemOpen
  \bibfield  {author} {\bibinfo {author} {\bibfnamefont {R.}~\bibnamefont
  {{Penrose}}}\ and\ \bibinfo {author} {\bibfnamefont {W.}~\bibnamefont
  {{Rindler}}},\ }\href@noop {} {\emph {\bibinfo {title} {{Spinors and
  space-time. Vol. 1: Two-spinor calculus and relativistic fields.}}}}\
  (\bibinfo {year} {1984})\BibitemShut {NoStop}%
\bibitem [{\citenamefont {{Blackburn}}\ \emph {et~al.}(2020)\citenamefont
  {{Blackburn}}, \citenamefont {{Pesce}}, \citenamefont {{Johnson}},
  \citenamefont {{Wielgus}}, \citenamefont {{Chael}}, \citenamefont
  {{Christian}},\ and\ \citenamefont {{Doeleman}}}]{Blackburn+2020}%
  \BibitemOpen
  \bibfield  {author} {\bibinfo {author} {\bibfnamefont {L.}~\bibnamefont
  {{Blackburn}}}, \bibinfo {author} {\bibfnamefont {D.~W.}\ \bibnamefont
  {{Pesce}}}, \bibinfo {author} {\bibfnamefont {M.~D.}\ \bibnamefont
  {{Johnson}}}, \bibinfo {author} {\bibfnamefont {M.}~\bibnamefont
  {{Wielgus}}}, \bibinfo {author} {\bibfnamefont {A.~A.}\ \bibnamefont
  {{Chael}}}, \bibinfo {author} {\bibfnamefont {P.}~\bibnamefont
  {{Christian}}},\ and\ \bibinfo {author} {\bibfnamefont {S.~S.}\ \bibnamefont
  {{Doeleman}}},\ }\bibfield  {title} {\bibinfo {title} {{Closure Statistics in
  Interferometric Data}},\ }\href {https://doi.org/10.3847/1538-4357/ab8469}
  {\bibfield  {journal} {\bibinfo  {journal} {\apj}\ }\textbf {\bibinfo
  {volume} {894}},\ \bibinfo {eid} {31} (\bibinfo {year} {2020})},\ \Eprint
  {https://arxiv.org/abs/1910.02062} {arXiv:1910.02062 [astro-ph.IM]}
  \BibitemShut {NoStop}%
\bibitem [{\citenamefont {{Wieringa}}(1992)}]{Wieringa1992}%
  \BibitemOpen
  \bibfield  {author} {\bibinfo {author} {\bibfnamefont {M.~H.}\ \bibnamefont
  {{Wieringa}}},\ }\bibfield  {title} {\bibinfo {title} {{An investigation of
  the telescope based calibration methods `redundancy' and `self-cal'}},\
  }\href {https://doi.org/10.1007/BF00420576} {\bibfield  {journal} {\bibinfo
  {journal} {Experimental Astronomy}\ }\textbf {\bibinfo {volume} {2}},\
  \bibinfo {pages} {203} (\bibinfo {year} {1992})}\BibitemShut {NoStop}%
\bibitem [{\citenamefont {{Liu}}\ \emph {et~al.}(2010)\citenamefont {{Liu}},
  \citenamefont {{Tegmark}}, \citenamefont {{Morrison}}, \citenamefont
  {{Lutomirski}},\ and\ \citenamefont {{Zaldarriaga}}}]{Liu+2010}%
  \BibitemOpen
  \bibfield  {author} {\bibinfo {author} {\bibfnamefont {A.}~\bibnamefont
  {{Liu}}}, \bibinfo {author} {\bibfnamefont {M.}~\bibnamefont {{Tegmark}}},
  \bibinfo {author} {\bibfnamefont {S.}~\bibnamefont {{Morrison}}}, \bibinfo
  {author} {\bibfnamefont {A.}~\bibnamefont {{Lutomirski}}},\ and\ \bibinfo
  {author} {\bibfnamefont {M.}~\bibnamefont {{Zaldarriaga}}},\ }\bibfield
  {title} {\bibinfo {title} {{Precision calibration of radio interferometers
  using redundant baselines}},\ }\href
  {https://doi.org/10.1111/j.1365-2966.2010.17174.x} {\bibfield  {journal}
  {\bibinfo  {journal} {\mnras}\ }\textbf {\bibinfo {volume} {408}},\ \bibinfo
  {pages} {1029} (\bibinfo {year} {2010})},\ \Eprint
  {https://arxiv.org/abs/1001.5268} {arXiv:1001.5268 [astro-ph.IM]}
  \BibitemShut {NoStop}%
\bibitem [{\citenamefont {Komesaroff}\ and\ \citenamefont
  {Lerchet}(1979)}]{Komesaroff+1979}%
  \BibitemOpen
  \bibfield  {author} {\bibinfo {author} {\bibfnamefont {M.~M.}\ \bibnamefont
  {Komesaroff}}\ and\ \bibinfo {author} {\bibfnamefont {I.}~\bibnamefont
  {Lerchet}},\ }\bibfield  {title} {\bibinfo {title} {Extending the fourier
  transform --- the positivity constraint},\ }in\ \href@noop {} {\emph
  {\bibinfo {booktitle} {Image Formation from Coherence Functions in
  Astronomy}}},\ \bibinfo {editor} {edited by\ \bibinfo {editor} {\bibfnamefont
  {C.}~\bibnamefont {Van~Schooneveld}}}\ (\bibinfo  {publisher} {Springer
  Netherlands},\ \bibinfo {address} {Dordrecht},\ \bibinfo {year} {1979})\ pp.\
  \bibinfo {pages} {241--247}\BibitemShut {NoStop}%
\bibitem [{\citenamefont {{Komesaroff}}\ \emph {et~al.}(1981)\citenamefont
  {{Komesaroff}}, \citenamefont {{Narayan}},\ and\ \citenamefont
  {{Nityananda}}}]{Komesaroff+1981}%
  \BibitemOpen
  \bibfield  {author} {\bibinfo {author} {\bibfnamefont {M.~M.}\ \bibnamefont
  {{Komesaroff}}}, \bibinfo {author} {\bibfnamefont {R.}~\bibnamefont
  {{Narayan}}},\ and\ \bibinfo {author} {\bibfnamefont {R.}~\bibnamefont
  {{Nityananda}}},\ }\bibfield  {title} {\bibinfo {title} {{The Maximum Entropy
  Method of Image Restoration Properties and Limitations}},\ }\href@noop {}
  {\bibfield  {journal} {\bibinfo  {journal} {\aap}\ }\textbf {\bibinfo
  {volume} {93}},\ \bibinfo {pages} {269} (\bibinfo {year} {1981})}\BibitemShut
  {NoStop}%
\bibitem [{\citenamefont {{Samuel}}\ and\ \citenamefont
  {{Nityananda}}(2000)}]{Samuel+2000}%
  \BibitemOpen
  \bibfield  {author} {\bibinfo {author} {\bibfnamefont {J.}~\bibnamefont
  {{Samuel}}}\ and\ \bibinfo {author} {\bibfnamefont {R.}~\bibnamefont
  {{Nityananda}}},\ }\bibfield  {title} {\bibinfo {title} {{Transport along
  null curves}},\ }\href {https://doi.org/10.1088/0305-4470/33/14/318}
  {\bibfield  {journal} {\bibinfo  {journal} {Journal of Physics A Mathematical
  General}\ }\textbf {\bibinfo {volume} {33}},\ \bibinfo {pages} {2895}
  (\bibinfo {year} {2000})},\ \Eprint {https://arxiv.org/abs/gr-qc/0005096}
  {arXiv:gr-qc/0005096 [gr-qc]} \BibitemShut {NoStop}%
\end{thebibliography}
%

\appendix

\section{Auto-correlations from Short-spaced Elements}\label{sec:short-spacing}

We have seen that auto-correlations can be used to construct covariants, not just advariants, from a triangle of elements such as in Eq.~(\ref{eqn:auto-corr-covariant}). This option is usually unavailable because of systematic errors in measuring auto-correlations. However, when the angular size of the object ($\theta_\textrm{obj}$) is small compared to the angular resolution determined by the element spacing, we can recover this advantage, by using a cross-correlation, $C_{00^\prime}$, when the elements $0$ and $0^\prime$ have a small separation ($D_{00^\prime}$), such that $\theta_\textrm{obj}\ll\lambda/D_{00^\prime}$, where, $\lambda$ is the wavelength of observation. 

We assume that the gains of the these closely spaced elements are independent of each other, while their correlation in $C_{00^\prime}$ is free from the problems that a single-element auto-correlation poses. One example would be an additional element, $0^\prime$, in close proximity to the base element 0. Another possibility is to use two physically close but independent subarrays (denoted by 0 and $0^\prime)$ that are phased from a dense array that is being used as a single element in a ``phased array'' mode. An example is the Atacama Large Millimeter/submillimeter Array (ALMA) in the EHT observations of M87 and Centaurus~A \cite{eht19-2,Janssen+2021}. Although $0^\prime$ can be paired with all the other existing elements, the true correlations, $S_{a0^\prime}$, will carry no new information that is not redundant\footnote{The redundancy, however, may be useful towards obtaining a calibration that is independent of knowledge of the sky brightness distribution, and forms the basis of redundant-calibration schemes in radio interferometry \cite[][for example]{Wieringa1992,Liu+2010}} with $S_{a0}$, excepting $S_{00^\prime}$, which is a good approximation to $S_{00}$.

Thus, we have two closely spaced elements, 0 and $0^\prime$, which do not resolve the object's features but have independent element gains, $G_0$ and $G_{0^\prime}$. This means that $S_{00^\prime}\approx S_{00}=S_{0^\prime 0^\prime}$. 
We can form the covariant, $C_{00^\prime} \widehat{C}_{0^\prime a} C_{ab} \widehat{C}_{b0}=G_0 S_{00^\prime} \widehat{S}_{0^\prime a} S_{ab} \widehat{S}_{b0} G_0^{-1} \approx S_{00} \widehat{S}_{0a} S_{ab} \widehat{S}_{b0}$. Note that the $G_0$ terms are eliminated because our group is Abelian and we have used $S_{00^\prime}\approx S_{00}$, which is the true auto-correlation. Thus, we have effectively included auto-correlations as the coincidence limit of cross-correlations, which will increase the number of real-valued independent invariants by 1 corresponding to the auto-correlation. The use of closely located elements was suggested in \cite{Broderick+2020} as a diagnostic for non element-dependent errors. Here, we are using them to provide one more invariant effectively involving an auto-correlation. 
 
Since we have used advariants as the building blocks, we can also reformulate this in terms of the advariant $C_{00^\prime} \widehat{C}_{0^\prime a}C_{a0}=G_0 S_{00^\prime} \widehat{S}_{0^\prime a} S_{a0}G_0^\dagger$, where, $a\notin \{0, 0^\prime\}$. Under our assumption of closeness of elements 0 and $0^\prime$, this can be written as $\approx G_0 S_{00} \widehat{S}_{0a}S_{a0}G_0^\dagger$. Since $\widehat{S}_{0a}S_{a0}=1$, we are therefore left with only $G_0 S_{00} G_0^\dagger$, which is exactly what an auto-correlation advariant pinned at 0 would have yielded. This means the earlier discussion with $n_A=1$ is applicable and we gain one invariant. This agrees with the preceding discussion based on covariants, as expected. 
 
We can also explain this using the dimension counting principles. The presence of an additional $S_{00^\prime}\approx S_{00}$, which is nearly real and positive, effectively increases the dimensionality of the real values in the true correlations by $n_\textrm{A}=1$ to $N(N-1)+1$. It would appear that we have introduced an unknown complex gain, $G_{0^\prime}$, consisting of two real parameters in the process. However, because of the redundancy $S_{a0^\prime}\approx S_{a0}$, $G_{0^\prime}$ can be expressed as
\begin{align}
  G_{0^\prime} &\approx C_{0^\prime a} C_{0a}^{-1} G_0 ,
\end{align}
which is fully determined by $G_0$, and therefore not an independent degree of gauge freedom. Hence, the number of unknown real parameters in the gains is still $2N-1$. Thus, the resulting number of real invariants increases by $n_\textrm{A}=1$, to $N^2-3N+2$, which confirms the alternate viewpoints presented above.

\section{Relation to self-calibration}\label{sec:selfcal}

The parameters describing the element-dependent effects can be determined if there is a standard signal. In practice, measurements  on a point-like object are interspersed with those of the target object, to get a preliminary calibration. A major improvement of this procedure is `self-calibration' \cite{Cornwell+1981,Pearson+1984,Ekers1984}. The approximate calibration parameters are only used as an initial guess to produce an approximate image. The image is then refined by alternating steps of deconvolution, with adjustment of the instrumental parameters to best fit the current image at each stage. This converges, in favourable cases, to a much better image than was earlier possible. The word `better' reveals that criteria based on \textit{a priori} information, such as positivity, smoothness, and compactness of the emission, play a role via the deconvolution step.    

Our work is concerned with a related but distinct notion of using closure invariants, which directly characterize the source and are independent of any model or deconvolution scheme. Forming images using just these invariants alone has been explored, especially in VLBI, by `forward-modeling', that is fitting a model to the measured closure invariants. This approach would be appropriate in cases where the data are not extensive enough to constrain a free-form fitting procedure like self-calibration. Invariants can be used to discriminate between different proposed models, purely in the domain of observations, without bringing in deconvolution with its attendant \textit{a priori} assumptions. A fully converged self-calibration  solution will, of course, automatically satisfy all invariants. However, invariants have a role even when self-calibration is not directly applicable.  

\section{Numerical Test for Independence of Invariants} \label{sec:numerics}

Here, we numerically verify the independence of the invariants derived through various analytical methodologies described in this paper. We begin with the simplest case with three elements which gives three cross-correlations. We also include one auto-correlation. 
We generate multiple realizations of one random positive real number for the uncorrupted auto-correlation, $S_{00}$, and three random complex numbers for the uncorrupted cross-correlations, $S_{01}$, $S_{12}$, and $S_{31}$, and three more random complex numbers for the element-based gains (corruptions), $G_0$, $G_1$, and $G_2$. These can be used to construct the measured correlations, $A_{00}$, $C_{01}$, $C_{12}$, and $C_{20}$ using Eq.~(\ref{mtrelation}), which are described by 7 real values. From these, we construct the advariants, $\mathcal{A}_{\Delta_1}$ and $\mathcal{A}_0$ and compute the triangular invariant, $\mathcal{A}_{\Delta_1}\widehat{\mathcal{A}}_0$, whose real and imaginary parts are the two closure invariants. 

Analytically, the test for independence of these invariants would be to look at the two real invariants as functions of the real and imaginary parts of the input true correlations. We construct the Jacobian matrix of partial derivatives relating first order changes in the output to those in the input. The rank of the Jacobian matrix gives the number of independent invariants. Numerically, we construct the elements of this Jacobian matrix by varying one input quantity $x_i$ at a time by a small amount, and recording the output changes $y_j$ as a column. In this case, there will be 7 columns of length 2, so the Jacobian is a $2 \times 7 $ matrix. The rank is checked numerically by carrying out a singular value decomposition (SVD) of the Jacobian and examining the list of non-zero singular values. Zero-valued singular values appear as very small values (compared to the non-zero singular values by many orders of magnitude) due to finite-precision computations. The number of zero singular values (also called the corank) gives the reduction in the number of independent invariants, compared to the total number being calculated. In geometric terms, we are finding the dimension of the surface onto which a given, general set of measured visibilities gets mapped when we compute a number of invariants of our own choosing. This dimension is in general less than the maximum possible rank, by the number of zeros. 

The rank of a Jacobian can also be used to cross-check the dimension counting. While we have pointed out the most obvious redundancy in the gains, namely, an overall phase, one might want to verify that no others have been missed. In this case, we simulate random true correlations as the input, and apply randomly chosen gains to them to get the measured correlations as output. Now, we vary all the gain parameters in small increments and find the Jacobian via the partial derivatives of all the real-valued parameters in the measured correlation with respect to these changes in the $2N$ real gain parameters. Not surprisingly, in all cases the corank is 1, and the count of $2N-1$ for the number of independent variations of the gains which modify the measured correlations, is confirmed. This numerical scheme generalizes to full polarimetric measurements as well.

\section{Choice of Invariants, Noise, and Imaging}\label{sec:invariants-form}

We have seen in section~\ref{sec:closure-invariants} that a complete and independent set of invariants is not necessarily unique and can take multiple forms. How do we choose the form for a complete and independent set of invariants from all possible sets? We approach this question from two considerations -- one from that of likelihood on the inferred model parameter space, and another from anticipated noise properties of the invariants themselves.

Let us consider determining the likelihood of a parametrized set of models, e.g., a ring of emission with azimuthal asymmetries as in \cite{eht19-6}. Here and below, we use a single symbol like $\mathcal{M}$ to denote an entire set of variables, in this case parameters describing the model. This model will lead to a predicted set of correlations from which a predicted set of  invariants follows, $\mathcal{I}^p(\mathcal{M})$. The noise on the correlations, presumed known, can be propagated to a probability distribution for the invariants $\mathcal{I}$
around $\mathcal{I}^p(\mathcal{M})$, using the functional relation $\mathcal{I}=f(C)$.   We denote this density  by $P_\mathcal{I}(\mathcal{I})$. Evaluating this at the measured values of the invariants $\mathcal{I}^m$ gives the likelihood function on the model space, $L(\mathcal{M}|\mathcal{I}^m )= P_\mathcal{I}(\mathcal{I}^m)$.

Now consider  working with two different, but  complete, sets of invariants, $\mathcal{I}$ and $\mathcal{\tilde{I}}$. There is a two-way functional relation between them. Because the probability densities for $\mathcal{I}$ and $\widetilde{\mathcal{I}}$ are related by $P_\mathcal{I}(\mathcal{I})\,\mathrm{d}\mathcal{I} = P_{\widetilde{\mathcal{I}}}(\widetilde{\mathcal{I}})\,\mathrm{d}\widetilde{\mathcal{I}}$, we get 
$L(\mathcal{M}|\mathcal{I}^m)=L(\mathcal{M}|\widetilde{\mathcal{I}}^m)\,|\mathbf{J}|$, 
where, $|\mathbf{J}|\coloneqq\det(\mathbf{J})$ is the determinant of the Jacobian matrix, $\mathbf{J}\equiv J_{pq}=\partial\widetilde{\mathcal{I}}_p/\partial\mathcal{I}_q$. $|\mathbf{J}|$ is evaluated at the \textit{measured} values of the two sets of invariants, and does \textit{not} depend on the model, $\mathcal{M}$. Hence, $|\mathbf{J}|$ appears as a simple proportionality factor in the space of $\mathcal{M}$. Therefore, the maximum likelihood solution does not depend on the choice of invariants.

In Bayesian approaches, including the maximum entropy methods, there is an additional factor, namely, a prior depending solely on $\mathcal{M}$. It may be noted that the standard approach of self-calibration does not even use invariants explicitly, while satisfying them implicitly,  so the question of a choice does not arise. It is therefore satisfying that direct use of invariants in determining the maximum likelihood of the models is also independent of the chosen form of invariants.

While the maximum likelihood of the models does not depend on the chosen form of the invariants, there may however be other considerations of noise characteristics and interpretations that may favour one form of invariants over another. Here, we compare and contrast our approach to that of \cite{Blackburn+2020}. Their construction of $N(N-3)/2$ independent closure amplitudes, and the demonstration that all other such amplitudes can be constructed from these is elegant and complete, and has the convenience of interpretation in terms of amplitudes and phases.

In our approach based on gauge theory, the use of triangles as the generators of all other loops and their associated invariants is more natural. These emerge as complex quantities, the magnitudes being  generalized closure amplitudes (since some have six baselines) and the phases are sums  of the standard  closure phases on two triangles. We have chosen to work with the real and imaginary parts rather than the amplitudes and phases of our set of invariants. Our primary motivation is described below.

Following \cite{TMS2017}, for simplicity, consider an ideal case of bivariate Gaussian distribution of two uncorrelated variables, $X$ and $Y$, denoting the real and imaginary parts of a complex number, respectively. Their joint probability distribution, without loss of generality, is taken to be centered on $(X_0,0)$ with a variance of $\sigma^2$, and is given by
\begin{align}
    P_{X,Y}(X,Y) &= \frac{1}{2\pi\sigma^2}\exp{\left[-\frac{(X-X_0)^2 +Y^2}{2 \sigma^2}\right]} \, . \label{eqn:cartesian-form}
\end{align}
Here, $P_{X,Y}(X,Y)=P_X(X)P_Y(Y)$. By change of variables, $X=A\cos\theta$ and $Y=A\sin\theta$, the joint probability distribution in polar form becomes
\begin{widetext}
\begin{align}
    P_{A,\theta}(A,\theta) &= \frac{A^\prime}{2\pi\sigma^2} \exp{\left[-\frac{(A-X_0)^2 - 2X_0^2\cos\theta + 4(A-X_0)X_0\sin^2\frac{\theta}{2}}{2 \sigma^2}\right]} \, . \label{eqn:polar-form}
\end{align}
\end{widetext}
where, $A^\prime=A\exp{\left[-X_0^2/\sigma^2\right]}$. The first and the second terms inside the exponential in Eq.~(\ref{eqn:polar-form}) depend only on $A-X_0$ and $\theta$, respectively. However, the third term depends on both, which makes $A$ and $\theta$ correlated in general. Only when $S/N \gg 1$ ($|A-X_0|\ll |X_0|$), the apparent correlation term may be neglected and the joint distribution becomes separable into amplitude and phase terms, $P_{A,\theta}(A,\theta)=P_A(A)P_\theta(\theta)$, where $P_\theta(\theta)$ reduces to a von Mises distribution, that can be further approximated as a Gaussian distribution in $\theta$ \cite{TMS2017,Blackburn+2020}.  
However, when $S/N \lesssim 1$, approximating that $A$ and $\theta$ are uncorrelated is not only invalid, but the phase is also not well-defined. So, the joint distribution is preferably represented in real and imaginary coordinates ($X$ and $Y$) rather than their polar form ($A$ and $\theta$). Thus, it is evident that purely on account of the choice of the coordinate system on which the joint probability distribution is represented, it can not only induce covariance among the random variables but also cause one or more variables to be poorly defined. Therefore, in such cases, a different coordinate system is clearly preferred. 

The closure invariants, regardless of the form they are represented in, are generally higher order functions of the correlations, and are thus expected to be neither Gaussian distributed nor have an uncorrelated behavior. When expressed as amplitudes and phases, they could suffer from induced covariance and singularities in low $S/N$ regimes. Thus, owing to such scenarios, we consider the real and imaginary parts of the covariants as ``better'' variables and prefer them over their amplitude and phase representation, even though they carry the same physical information.


\section{Connection to Pancharatnam phases}\label{sec:Pancharatnam-closure}

The electric field time-series, $e_a(t)$, received at element, $a$, as a function of time can be viewed as a vector in Hilbert space  \cite{Komesaroff+1979,Komesaroff+1981},
\begin{align}
    \ket{e_a} &= \{e_a(t)\,|\,0\le t\le T\} \, , \label{eqn:easavector}
\end{align}
where, $T$ is the integration time. The correlation function, $\left\langle e_a e_b^\dagger \right\rangle$, provides
an inner product between these vectors,
\begin{align}
    \braket{e_b|e_a} &= C_{ab} = \frac{1}{T}\int_0^T e_b^*(t) \, e_a(t) \, \mathrm{d}t \, . \label{eqn:corrasinnerproduct}
\end{align}
Closure phases can now be understood as a Pancharatnam phase \cite{Pancharatnam1956a,Pancharatnam1975} (see \cite{bhatnagar+2001} for an earlier radio astronomy application). Pancharatnam's work in polarization optics gives us a rule (known in mathematics as a ``connection'') for comparing phases between vectors based at $a$ and $b$. We fix $\ket{e_a}$ at element $a$ and define $\ket{e_b}$ to be {\it in phase}
with $\ket{e_a}$ if the correlation  $\braket{e_b|e_a}$ is real and positive. The physical motivation is  that the intensity of the superposed beam achieves a maximum under this condition. This gives 
a rule for transporting a phase from element $a$ to $b$, that is, modifying the phase of $\ket{e_b}$ so that it is ``in agreement'' with $\ket{e_a}$. Iterating this
rule and going in a closed loop from element $a$ to $b$ to $c$ and back
to $a$, we find on returning to $a$ that the cyclic application of the rule gives a non-trivial phase change with respect to the original phase of $\ket{e_a}$. In words, $\ket{e_b}$ can be ``in phase'' with $\ket{e_a}$, and $\ket{e_c}$ with $\ket{e_b}$, but then $\ket{e_c}$ in general is not ``in phase'' with $\ket{e_a}$. The Pancharatnam phase is 
\begin{align}
    \arg \braket{e_b|e_a}\braket{e_c|e_b}\braket{e_a|e_c} &= \arg C_{ab} C_{bc} C_{ca} \, , \label{eqn:panch}
\end{align}
which astronomers will recognize as the closure phase \cite{jen58}.
Closure phases, thus, emerge as the curvature of the Pancharatnam rule 
for comparing phases. This is a discrete version of the curvature
familiar from parallel transport along a closed curve on a sphere.

A natural question arises at this point: is there a transport rule for the \textit{amplitude} as well as the phase? The idea is to modify the vector $\ket{e_b}$ representing the signal at $b$ so that it is `in agreement' with $\ket{e_a}$ in both phase and amplitude. This is achieved as follows. The vector $\ket {e_b}' =f_b \ket{e_b}$ is proportional to $\ket {e_b}$, but is now rescaled by a complex  factor  $f_b$, i.e., in amplitude and phase so that $\braket{e_a|e_b'}=f_b \braket{e_a|e_b}=1$ (note that in the case of phases this was only made real and positive, but the magnitude was left undetermined). This gives $f_b=1/\braket{e_a|e_b}$. At the next stage, the vector $\ket{e_c}$ is rescaled to $f_c \ket{e_c}$ so that its inner product with $f_b\ket{e_b}$ is 1, i.e., $ f_b^* f_c \braket{e_b|e_c}=1$. This gives $f_c=\braket{e_a|e_b}^*/\braket{e_b|e_c}$. The pattern is now clear. After an even number of steps, returning to $a$, we obtain a vector $f_a \ket{e_a}$ which can be compared in amplitude and phase to the original $\ket{e_a}$. The rescaling factor, $f_a$, comes out to be nothing but our four element complex closure invariant, made up of correlations with alternate terms hatted. This transport rule, applicable to amplitudes and phases, can be regarded as a spinoff from radio astronomy to possible application in other areas. The restriction to an even number of steps has appeared before in relativity in a discussion of the analogue of Fermi transport for null curves \cite{Samuel+2000}.

\end{document}